\documentclass[numbers,webpdf,imaiai]{ima-authoring-template}%

\graphicspath{{Fig/}}

\theoremstyle{thmstyletwo}%

\usepackage{natbib}
\setcitestyle{round}

\numberwithin{equation}{section}
\makeatletter
\renewcommand\@biblabel[1]{}
\makeatother

\begin{document}

\DOI{}
\copyrightyear{2021}
\vol{00}
\pubyear{2021}
\firstpage{1}


\title[Sectoral PO by judicious selection of financial ratios via PCA]{Sectoral portfolio optimization by judicious selection of financial ratios via PCA}

\author{VRINDA DHINGRA \ORCID{0000-0002-1035-5097}
\address{\orgdiv{Department of Mathematics}, \orgname{Indian Institute of Technology, Roorkee}, \orgaddress{\postcode{247667}, \state{Uttarakhand}, \country{India}}}} 
\author{AMITA SHARMA \ORCID{0000-0003-3933-3999}
\address{\orgdiv{Department of Mathematics}, \orgname{Netaji Subhas University of Technology}, \orgaddress{\postcode{110078}, \state{New Delhi}, \country{India}}}}
\author{SHIV KUMAR GUPTA* \ORCID{0000-0003-2613-6806}
\address{\orgdiv{Department of Mathematics}, \orgname{Indian Institute of Technology, Roorkee}, \orgaddress{\postcode{247667}, \state{Uttarakhand}, \country{India}}}}

\authormark{Vrinda Dhingra et al.}

\corresp[*]{Corresponding author: \href{email:email-id.com}{s.gupta@ma.iitr.ac.in}}

\received{Date}{0}{Year}
\revised{Date}{0}{Year}
\accepted{Date}{0}{Year}

\editor{Associate Editor: Name}

\abstract{Embedding value investment in portfolio optimization models has always been a challenge. In this paper, we attempt to incorporate it by employing principal component analysis to filter out dominant financial ratios from each sector and thereafter, use the portfolio optimization model incorporating second order stochastic dominance criteria to derive an optimal investment. We consider a total of $11$ financial ratios corresponding to each sector representing four categories of ratios, namely liquidity, solvency, profitability, and valuation. PCA is then applied over a period of 10 years to extract dominant ratios from each sector in two ways, one from the component solution and other from each category on the basis of their communalities. The two step Sectoral Portfolio Optimization (SPO) model is then utilized to build an optimal portfolio. The strategy formed using the former extracted ratios is termed as PCA-SPO(A) and the latter as PCA-SPO(B). The results obtained from the proposed strategies are compared with those from mean variance, minimum variance, the SPO and the nominal SSD models, with and without financial ratios. Empirical performance of proposed strategies is analysed in two ways, viz., using a rolling window scheme and using market trend scenarios for S\&P BSE 500 (India) and S\&P 500 (U.S.) markets. We observe that the proposed strategy PCA-SPO(B) outperforms all other models in terms of downside deviation, CVaR, VaR, Sortino, Rachev and STARR ratios over almost all out-of-sample periods. This highlights the importance of value investment where ratios are carefully selected and embedded quantitatively in portfolio selection process.}

\keywords{Financial Ratios; In-sample and out-of-sample analysis; Principal component analysis; Second order stochastic dominance; Sectoral portfolio optimization}

\maketitle

\bigskip

\section{Introduction}
\label{intro}
Benjamin Graham (\citeauthor{grahamdodd} \citeyear{grahamdodd}; \citeauthor{graham} \citeyear{graham}) and Warren Buffet are among the famous personalities who emphasized on the importance of using fundamental analysis into investment, also known as value investment. The aim of fundamental analysis is to pick fundamentally strong assets to produce robust performance of a portfolio over the investment period. However, when and how to include the FRs quantitatively in portfolio selection process impacts its performance significantly. 

Financial analysis (FA) which is a subset of fundamental analysis, focuses on financial ratios (FRs) to identify the strengths and weaknesses of a company in terms of its financial performance e.g. sales, earnings, debt, liquidity, and liabilities among others. Financial ratio (FR) is the ratio of two appropriately chosen numerical values of the financial entities of a company to figure out the relative performance. For example, current ratio collates current assets to current liabilities, gauging a firm's ability to meet its short term obligations whereas net profit margin is the net profit earned by a company as a percentage of its total revenue, highlighting overall growth of the firm. The broad categories into which the FRs can be classified are: profitability, liquidity, solvency, and valuation.  

The importance and usefulness of FRs has been repeatedly demonstrated in various studies (\citeauthor{beaver} \citeyear{beaver}; \citeauthor{chenshirmeda} \citeyear{chenshirmeda}; \citeauthor{delen2013} \citeyear{delen2013}; \citeauthor{arkan} \citeyear{arkan}). Researchers, accountants, and financial analysts have been using financial ratios for the assessment of risks, forecasting failures, predicting future profits, and liquidity positions among others. The importance of FRs is not limited to forecasting and predictive analysis, it has also been adopted in portfolio selection. For instance, several researchers (\citeauthor{edrizhang2007} \citeyear{edrizhang2007}; \citeauthor{samaras} \citeyear{samaras}) have incorporated the FRs by employing multi-criteria decision making techniques to filter out elite stocks and then construct an optimal portfolio. On the other hand, few others (\citeauthor{sharmamehra_spo} \citeyear{sharmamehra_spo}; \citeauthor{tarczynski} \citeyear{tarczynski}) emphasized on the importance of sector-wise comparison of the assets when FRs is one of the selection criteria. 

It is well-known that the sectors behave differently owing to their varied business functionalities and operational structures and therefore, comparison between two companies belonging to different sectors\footnote{In financial markets, sectors are the groups of stocks classified according to their similar business functionality and operations} on the basis of same FRs is not wise. Very few studies focus on this important step. The authors (\citeauthor{sharmamehra_spo} \citeyear{sharmamehra_spo}) attempted to resolve this issue by proposing a two-step investment scheme, termed as SPO in their paper, by first solving the second order stochastic dominance (SSD) portfolio optimization model for each sector separately on the basis of their FR values and then a final optimal portfolio is obtained by combining the weights derived from step 1 along with their mean return performance. They used the SSD criteria from the benchmark index in constraints to generate portfolios in both the steps. 

SPO, though having several advantages over the classical SSD model, incorporates value investment in the form of four fixed FRs common for all sectors. However, it is well known that ratios that are found to be critical for one sector may not be so for another sector because companies in different sectors have different market and capital structures. For instance, the banking and finance sector rely on credit calculations, therefore, liquidity and solvency ratios are more important to this sector whereas the construction industry being capital intensive relies hugely on equipment to build projects and thus profitability ratios are most valuable among construction-based firms. Hence, the choice of ratio(s) in each sector is crucial in the decision making process and thus considering fixed set of FRs for all sectors is unable to reflect the importance of value investment. 

To address this issue, we propose to apply Principal Component Analysis (PCA) sector wise to extract most dominant financial ratios for each sector and thereafter, use the SPO two step scheme to generate final optimal portfolio. PCA is a non-parametric statistical tool that transforms a set of variables into few factors called principal components, retaining maximum information of the original variables. We offer to extract four dominant ratios explaining atleast 80\% of the total variation in each sector from PCA in two ways, first by fetching the four ratios corresponding to each principal component and second by selecting from each of the four categories of ratios on the basis of their communalities. We then use the SPO model to derive final optimal portfolios constructed from both types of extractions. We named the proposed strategy PCA-SPO(A) in case of deriving the overall four dominant ratios and PCA-SPO(B) when each ratio becomes the dominant ratio in its category. 

To gauge the importance of sector-wise usage of value investment in portfolio selection, we also solve the classical SSD (\citeauthor{dentruzy03} \citeyear{dentruzy03}) model with FRs (named as F-SSD) and without FRs (named as SSD). While the SSD model considers only mean return values in its objective function, the F-SSD model combines the mean return values with the FRs in the objective function. PCA is being utilized when all assets are taken together to select the four dominant ratios in case of the F-SSD model. Also, both the SSD and F-SSD models are optimized on all assets at once. Therefore, an improvement of the proposed strategies over F-SSD or SSD reflects the importance of utilizing FRs sector wise in asset allocation. In addition to the SSD model, we also solve the \citeauthor{markowitz} \citeyearpar{markowitz} mean-variance model (named as Mean-var) and minimum variance (\citeauthor{JagannathanMa} \citeyear{JagannathanMa}) model (named as Min-var) to broaden the comparison analysis.

For the purpose of computational analysis, we consider historical data of the FRs and the closing prices for the stocks listed in the $6$ sectors, namely Information Technology (IT), Consumer Durable (CD), Energy, Banking and Finance, Fast Moving Consumer Goods (FMCG), and Healthcare of two markets, viz., S\&P BSE 500 of India and S\&P 500 of the United States(U.S.). We chose a set of total $11$ FRs representing four important categories, Profitability, Liquidity, Solvency and Valuation for each of these 6 sectors for a period of $10$ years from April 2004 to March 2014 to filter out dominant ratios using PCA. We then solve all the models under the in-sample and out-of-sample framework.


To make the analysis comprehensive, we carry out the out-of-sample analysis in two ways. First, we use a rolling window scheme of 4 weeks (or one month) over a period of of 6 years from April 2014 to March 2020, with an in-sample period of 52 weeks (or one year) and out-of-sample length as 13 weeks. Secondly, we analyze the performance of the portfolios from the models on three different out-of-sample market scenarios, namely, neutral, bearish and bullish. We solve a total of seven optimization models, Mean-var, Min-var, SPO, SSD, F-SSD and two models corresponding to the proposed strategies PCA-SPO(A) and PCA-SPO(B). The performance of the portfolios generated from these models is measured and analyzed on many fronts, including mean return, several risk measures and performance ratios. We consider, Conditional Value at risk (CVaR) at 95\% and 97\%, Downside Risk, Value at risk (VaR) at 95\% and 97\% as risk measures and Sharpe, Sortino, STARR and Rachev as performance measures.  

We observe that for the rolling window scheme, the portfolios obtained from the proposed strategies PCA-SPO(A) and PCA-SPO(B) depicts a better risk-reward profile than all other models by achieving higher values of Sortino, Sharpe, Rachev and STARR ratios and lower risk values in terms of DD, VaR and CVaR. On the other hand, for the phase-wise analysis, while PCA-SPO(B) proves favourable for moderate investors only during a bullish phase, it proves advantageous for all types of investors for the bearish and neutral phases. This highlights the benefit of considering ratios from each category in the process of portfolio construction. We also note down that the F-SSD model suffers with the highest risk and under performs in other measures in comparison to other models, indicating that mere consideration of financial ratios in portfolio selection may not be wise and its correct implementation in form of sector-wise comparison is even more crucial. 

To summarize, the major contributions of this study are: (i) firstly, it uses PCA to determine the appropriate objective functions, viz., the financial ratios that helps pick out the best stocks from each sector in terms of their financial strength, thereby overcoming the limitation of the SPO model. This is crucial from the point of value investment because each sector has its own business functionalities and market structure. Thus, the ratios that best determine the financial health of one sector may not be so for the other sector; (ii) Secondly, though this article can be seen an extension of the previous work of \citeauthor{sharmamehra_spo} \citeyearpar{sharmamehra_spo}, however, the contribution of this paper is rich in terms of its empirical analysis that is extensively carried out using the rolling window strategy and on different scenarios of the market.

The rest of the paper is structured as follows. Section 2 presents a brief review of literature on SSD, FA and PCA in portfolio selection. Section 3 describes the SPO model. Section 4 elucidates the proposed strategy. Section 5 communicates the sample data, PCA results, the in-sample and out-of-sample performance analysis. Section 6 outlines the managerial implications and Section 7 concludes the paper.

\vspace{-5pt}

\section{Literature Review}
\label{sec:2}
We now present a brief survey of literature on SSD, PCA and FA in portfolio selection process. 

\subsection{Second order Stochastic Dominance}
\label{sec: 2.1}
 Stochastic Dominance (SD) stems from the theory of majorization (\citeauthor{marshellolkin} \citeyear{marshellolkin}) that primarily ranks two real-valued random vectors. SD is popularly known for its relationship with the different classes of utility functions (\citeauthor{levy92} \citeyear{levy92}). Of utmost importance to economics are SD of order one (FSD; \citeauthor{quirksapo62} \citeyear{quirksapo62}), order two (SSD; \citeauthor{hadarrussell} \citeyear{hadarrussell}) and order three (TSD; \citeauthor{whitmore} \citeyear{whitmore}), for they respectively resonate the behaviour of the class of rational, rational risk-averse, and rational risk-averse and ruin-averse investors. Among the three orders, the SSD is widely used in portfolio optimization as it describes the preference of all rational, risk-averse investors, characterized by non-decreasing concave utility functions. 

The traditional mean-risk formulations are unable to extract much information from the return distribution, and thus some researchers recommended the incorporation of either more than one risk measures or moments (\citeauthor{konno93} \citeyear{konno93}; \citeauthor{konno95} \citeyear{konno95}; \citeauthor{amita1} \citeyear{amita1}; \citeauthor{omega_amita} \citeyear{omega_amita}; \citeauthor{zhaoet15} \citeyear{zhaoet15}) or SD (\citeauthor{romanmitra2009} \citeyear{romanmitra2009}) in portfolio optimization framework. Apart from being theoretically sound, portfolio optimization models incorporating SSD criteria from the benchmark portfolio are computational efficient (\citeauthor{fabianmitra2011} \citeyear{fabianmitra2011}; \citeauthor{bruni2012} \citeyear{bruni2012}; \citeauthor{romanmitra2013} \citeyear{romanmitra2013}; \citeauthor{sharmaagrmehra2017} \citeyear{sharmaagrmehra2017}; \citeauthor{singh2017dharamraja} \citeyear{singh2017dharamraja}; \citeauthor{sharmamehra_spo} \citeyear{sharmamehra_spo}; \citeauthor{goelsharma_ssddev} \citeyear{goelsharma_ssddev}; \citeauthor{sehgal2020robustssd} \citeyear{sehgal2020robustssd}).  \citeauthor{dentruzy03} \citeyearpar{dentruzy03, dentruzy06} used the lower partial moment (LPM) characterization of SSD in the constraints from the benchmark portfolio. \citeauthor{romanmitra2006} \citeyearpar{romanmitra2006} investigated the applications of SSD in enhanced indexing by implementing the cutting plane algorithm. \citeauthor{fabianmitra2011} \citeyearpar{fabianmitra2011} proposed a portfolio optimization model incorporating SSD using tail risk measures. \citeauthor{romanmitra2013} \citeyearpar{romanmitra2013} analyzed the applications of SSD in portfolio generation using the reference point method. \citeauthor{fidankececi} \citeyearpar{fidankececi} compared the SSD constrained model with minimum variance and mean-variance models and observed out-performance of the former in terms of high returns and Sharpe ratio. \citeauthor{sharmaagrmehra2017}\citeyearpar{sharmaagrmehra2017} relaxed the SSD condition by introducing under-achievement and over-achievement variables for the purpose of enhanced indexing in view of the almost SD introduced by \citeauthor{leshnolevy} \citeyearpar{leshnolevy}.

\citeauthor{lvetal20} \citeyearpar{lvetal20} proposed an incremental bundle algorithm using inexact oracle for solving SSD constrained portfolio optimization (PO) model. \citeauthor{guran} \citeyearpar{guran} applied the mean-variance PO on the energy sector stocks by first eliminating inefficient stocks with the help of SSD efficiency test. \citeauthor{goelsharma_ssddev} \citeyearpar{goelsharma_ssddev} introduced deviation measure in SSD and explored its application to enhanced indexing. \citeauthor{sehgal2020robustssd} \citeyearpar{sehgal2020robustssd} proposed a robust portfolio model with SSD constraints. Recently, \citeauthor{kopa_ima} \citeyearpar{kopa_ima} used four types of stochastic dominance relations including SSD for pension fund selection and \citeauthor{sectorSD_ima} \citeyearpar{sectorSD_ima} investigated the consistency of sector weighted portfolios with multivariate stochastic dominance.


 \subsection{Financial Analysis}
 \label{sec: 2.2}
 
 Considering the movement in share price alone is a naive way to evaluate the performance of an asset. It is therefore advisable to invest in a company which is doing good in terms of its financial entries or FRs. \citeauthor{edrizhang2007} \citeyearpar{edrizhang2007, edrizhang2008} employed 18 financial ratios to analyze a firm's financial strength by proposing a generalized Data Envelopment Analysis (DEA) model. They first selected stocks on the basis of RFSI (relative financial strength indicator) and then formed an optimal portfolio using the mean-variance model. The RFSI was calculated in two ways, one using the raw indicators (total assets, long term debt, accounts receivable, net income and total liabilities) from the financial statements and other using the financial ratios (profitability, liquidity, leverage, valuation and growth ratios). It was observed that RFSI calculated using the latter was shown to have higher correlation with stock returns and the portfolio thus generated performed better. This highlighted the importance of FA in portfolio selection. Some authors (\citeauthor{samaras} \citeyear{samaras}; \citeauthor{xidonas} \citeyear{xidonas}) used FA to first filter stocks and then used Multi-Criteria Decision Making (MCDM) techniques to construct optimal portfolios. \citeauthor{huang14} \citeyearpar{huang14} employed FA using three key indicators- gross profit margin, return on equity and cash flow growth rate to screen fundamentally strong stocks and then an optimal portfolio was generated with the help of integrated DEA-Multi-Objective Decision Making (DEA-MODM) model. \citeauthor{yuetal2009} \citeyearpar{yuetal2009} and \citeauthor{silva2014} \citeyearpar{silva2014} used several fundamental indicators like ROE (return on equity), Debt-ratio, and P/E ratio jointly with technical indicators to generate optimal portfolios using evolutionary algorithms. 
 
\citeauthor{tarczynski} \citeyearpar{tarczynski} emphasized on identifying the leading sectors in portfolio analysis. They employed several FRs, namely current ratio, return on assets, return on equity, and debt-equity ratio along with other indicators like size of sector to construct a fundamental power index (FPI) of sectors. \citeauthor{sharmamehra_spo} \citeyearpar{sharmamehra_spo} designed a two-step SSD portfolio optimization model where the stocks were optimized sector-wise on the basis of four FRs, namely return on assets, current ratio, debt-asset ratio, and P/E ratio to investigate the importance of sector wise selection. \citeauthor{jothimani17} \citeyearpar{jothimani17} formulated a PCA-DEA model for stock selection. They used several FRs as input and output parameters, which were first filtered using PCA and thereafter utilized to select efficient stocks by the DEA. \citeauthor{deng2021} \citeyearpar{deng2021} designed an intelligent system integrating PCA to identify insider tradings in Chinese security markets. Financial ratios such as P/E ratio, P/B ratio, ROA(return on asset), ROE(return on equity), QR(quick ratio), among others have been used as indicators for insider trading sample and as inputs for PCA. 


\subsection{Principal Component Analysis} 
\label{sec 2.3}
 
 Principal Component Analysis (PCA), a subset of Factor analysis is a statistical tool mainly used for dimensionality reduction in wide variety of domains like computer vision and image processing (\citeauthor{hernandezmendez} \citeyear{hernandezmendez}), drug discovery and biomedical data (\citeauthor{giuliani} \citeyear{giuliani}), and material sciences (\citeauthor{baydar_ima} \citeyear{baydar_ima}). In Finance, PCA is used for time series forecasting (\citeauthor{chowdhury2018} \citeyear{chowdhury2018}), financial risk ranking (\citeauthor{fangetal} \citeyear{fangetal}), forecasting stock prices (\citeauthor{wang&wang} \citeyear{wang&wang}; \citeauthor{zhong-enke} \citeyear{zhong-enke}; \citeauthor{ghorbani} \citeyear{ghorbani}), selection of relevant financial ratios for predictive analysis (\citeauthor{deng2021} \citeyear{deng2021}; \citeauthor{olufemietal2012} \citeyear{olufemietal2012}) and stock selection for optimal portfolio generation (\citeauthor{yuetal2014} \citeyear{yuetal2014}; \citeauthor{joldes} \citeyear{joldes}). 
 
 Factor analysis for financial ratios was first applied by \citeauthor{pinches73} \citeyearpar{pinches73} where they reduced their data set of bond ratings from 35 variables to 7 variables, retaining around $63\%$ of the total variation. \citeauthor{tanetal97} \citeyearpar{tanetal97} used factor analysis on 29 ratios for a period of 12 years and derived 8 underlying factors for companies listed in Singapore. \citeauthor{ocaletal} \citeyearpar{ocaletal} applied factor analysis on the Turkish construction industry and reduced 25 financial ratios to 5 significant factors. \citeauthor{benchinyap} \citeyearpar{benchinyap} investigated the application of PCA on 28 financial ratios for two industry sectors, namely consumer and trading sector and found that a smaller set of 7 and 9 ratios, respectively were sufficient and industry specific.
 
 \citeauthor{fulgadedu} \citeyearpar{fulgadedu} designed a three-step model wherein they first employ PCA and clustering techniques to generate classes of similar stocks. Stocks with minimal Value-at-Risk (VaR) in each class are then filtered and an optimal portfolio is constructed using mean-VaR model on these filtered stocks. \citeauthor{yuetal2014} \citeyearpar{yuetal2014} formulated a PCA-SVM (Support Vector Machine) stock selection model. They first implemented PCA to reduce the dimensionality of the training set that comprised of data of 20 financial ratios from 677 companies and then employed SVM algorithm for the selection of final stocks. \citeauthor{nadkarnineves} \citeyearpar{nadkarnineves} combined PCA with a neural network algorithm to generate optimal trading signals. 
 
We observed that either the researchers have totally ignored the importance of FA in the field of portfolio optimization or even if they did, then they implemented FA using different FRs depending on their popular usage. However, which ratio or a set of ratios one should use to analyze the objective at hand is a recurring question. Ideally, the selection of financial ratios should not merely be based on the popularity of their usage but rather on some theoretical or empirical evidence. The SPO model (\citeauthor{sharmamehra_spo} \citeyear{sharmamehra_spo}) effectively implements the use of FR sector-wise in the optimization model but uses fixed four ratios, common for all sectors. This paper is a very first attempt to improve the SPO model by implementing Principal Component Analysis (PCA) for the selection of dominant ratios for each sector.

\section{The SPO Model}
\label{sec:3}

\subsection{Financial Ratios and sectors}
\label{sec: 3.1}
To make the paper self-sufficient, we begin by first describing the financial ratios and the economic sectors relevant to the paper, followed by briefly sketching out the SPO Model. 

\begin{enumerate}
\item \textbf{Financial Ratios:} Benjamin Graham, regarded as the father of fundamental analysis, was the first to idealize the concept of financial ratio analysis (\citeauthor{grahamdodd} \citeyear{grahamdodd}; \citeauthor{graham} \citeyear{graham}). Financial ratio analysis, a subset of fundamental analysis, analyzes the financial statements of a company to determine its overall financial health. Financial Ratios can be broadly classified into four categories: liquidity, profitability, solvency/leverage, and valuation whose brief description is given below: 

\begin{enumerate}
    \item  \textit{Profitability Ratio (PR)} measures and evaluates the ability of a firm to utilize its resources to generate profit. Profits generated by a firm are used to fund future developments and pay off dividends to the shareholders. Therefore, the efficiency of a firm to make profits is an important factor to look after. A higher value of PR is favourable. Gross/net profit margin, return on assets (ROA), return on equity (ROE), earning per share (EPS) are some of the well known profitability ratios.

    \item \textit{Liquidity Ratio (LR)} indicates the ability of a firm to meet its short-term debt obligations. A higher value of LR is preferable. Current ratio, quick ratio, operating cash flow to current liabilities, are few examples of liquidity ratios.

    \item \textit{Solvency Ratio (SR)} depicts the financial risk involved in terms of long-term borrowings. More borrowings imply higher risk and thus a lower solvency ratio is desirable. Debt-equity/assets ratio, interest coverage ratio, capitalization ratio, are some well known solvency ratios.

    \item \textit{Valuation Ratio (VR)} helps to determine the relationship between the current share price of a company with its intrinsic value. It helps in estimating whether a company is exorbitant or cheaper compared to its growth, earnings and other prospects. Price-to-earning/sales (P/E), price-to-book ratio (P/B), dividend yield, and enterprise value multiple, are examples of valuation ratios.
    \end{enumerate}

\item \textbf{Sectors:} A sector is a large fragment of an economy where companies that are engaged in a similar businesses are involved, like Auto, Finance, and Healthcare among others. Based on market movement, the sectors are broadly classified as cyclical, defensive (non-cyclical), and sensitive. Cyclical and sensitive sectors are the ones that follow market trend, while cyclical sectors are highly sensitive to the market peaks and troughs, the sensitive sectors are comparatively moderate in their correlation to such movements. The defensive sectors, on the other hand are those which are least affected with the market oscillations. The defensive sectors are more stable in terms of their earnings than the cyclic and sensitive sectors whose returns are comparatively volatile. Therefore, while comparing stocks on the basis of FA, a firm's solvency in terms of debt-equity/assets ratio or valuation ratios like P/B and P/E ratios demonstrates better health of a cyclical or sensitive stock and a firm's earnings figures like return on asset/equity are more relevant for the defensive sectors. 
\end{enumerate}

Table \ref{tab:sectors} describes the major sectors of the economy along with their classification and composition in the S\&P BSE 500\footnote{https://www.bseindia.com/downloads1/spbsesectordiversificationconsultation5222019.pdf} and S\&P 500 markets\footnote{https://www.thebalance.com/what-is-the-sector-weighting-of-the-s-and-p-500-4579847}. While the finance sector dominates the S\&P BSE 500 index, the S\&P 500 index is heavily weighted towards information technology. 

\begin{table}[!htp]
\centering
\caption{Sector classification and composition in S$\&$P BSE 500 (India) and S$\&$P 500 (U.S.) indices}
\begin{tabular*}{\textwidth}{@{\extracolsep{\fill}}lllcc@{\extracolsep{\fill}}}
\toprule
 S.No.   & Sectors & Classification   & \% weight & \% weight in\\
         &     & & in S\&P 500 &  S\&P BSE 500     \\
\noalign{\smallskip}\hline\noalign{\smallskip}
1 & Finance &                   Cyclic &      11.5 &    33.5   \\
2 & Basic Materials  &          Cyclic &      2.60 &    7.03\\ 
3 & Information Technology (IT) &   Sensitive &  28.1 & 10.7 \\
4 & Communication Services  &  Sensitive &  9.60 & 1.41 \\
5 & Energy (Oil \& Gas) &           Sensitive &  3.70 & 9.77\\
6 & Consumer Durables (CD) &    Defensive &  4.50 & 3.00\\
7 & FMCG (Consumer Staples) &   Defensive & 6.20 & 9.32 \\
8 & HealthCare &            Defensive & 13.3 & 5.07\\
9 & Utilities &          Defensive &  2.60 & 3.00\\
10 & Industrials &      Defensive &  8.00 & 7.98\\
11 & Miscellaneous &    Mixed &  9.90  & 9.30\\
\hline\noalign{\smallskip}
\end{tabular*}
\label{tab:sectors}
\end{table}

We now describe the two step SPO model, which optimizes the portfolio on the basis of FR sector wise. The following notations are used throughout the paper.

\vspace{-2cm}

\subsection{Notations}
\label{sec:3.1}
\begin{align*}
       T \ \quad \qquad \qquad \qquad \qquad \qquad \qquad & \text{  total number of scenarios of the time horizon $\Gamma$} \\
       N \  \quad \qquad \qquad \qquad \qquad \qquad \qquad  & \text{  total number of assets} \\
       s  \ \ \quad \qquad \qquad \qquad \qquad \qquad \qquad  &  \text{  total number of sectors} \\
       n_r \  \quad \qquad \qquad \qquad \qquad \qquad \qquad  &  \text{  total number of assets in sector $r$} \\
       z=(z_{1}, z_{2}, \ldots, z_{N})'\in \mathbb{R}^{N} \  \quad \quad \qquad  & \text{  portfolio of $N$ assets; $z_{j}$ denotes fraction of } \\
        \quad \quad \quad \quad  \quad   \qquad \qquad \qquad \qquad \qquad  & \text{  total budget invested in asset}\; j\\
        z^{r}=(z_{1}^{r}, z_{2}^{r}, \ldots, z_{n_r}^r)' \quad  \quad \qquad \qquad  & \text{  portfolio weights vector for $r$th sector}\\
        r_{j} \ \quad \qquad \qquad \qquad \qquad \qquad \qquad   & \text{ random return from $j$th asset}\\
    r_{jt} \quad \qquad \qquad \qquad \qquad \qquad \qquad   & \text{ return realization from $j$th asset at scenario $t$}\\
    p_{t} \ \quad \qquad \qquad \qquad \qquad \qquad \qquad  & \text{  probability of scenario $t$}\\
    R_z  \quad \qquad \qquad \qquad \qquad \qquad \qquad  & \text{ random return of portfolio $z$}\\
    R_{zt}=\sum\limits_{j=1}^{N}r_{jt}z_{j}  \ \ \qquad \qquad \qquad \qquad  & \text{ return realization of portfolio $z$ at scenario $t$}\\
    E(R_{z})=\sum\limits_{t=1}^{T} \left(\sum\limits_{j=1}^{N}r_{jt}z_{j}\right)p_{t} \ \quad \qquad & \text{ expected value of $R_{z}$}\\
    R_{z^{r}}  \quad \qquad \qquad \qquad \qquad \qquad \qquad  & \text{ random return of $r$th sector portfolio $z^{r}$}\\
    Y \ \quad \qquad \qquad \qquad \qquad \qquad \qquad & \text{ random return of benchmark index } \\
    Y^{r}  \quad \qquad \qquad \qquad \qquad \qquad \qquad & \text{ random return of sectoral benchmark }\\
  \qquad   \qquad \qquad \qquad \qquad \qquad \quad & \text{ index for $r$th sector }
\end{align*}

 \subsection{Sector Portfolio Optimization Model (SPO)}
 \label{sec: 3.3}
 The SPO model is a two-step PO model (\citeauthor{sharmamehra_spo} \citeyear{sharmamehra_spo}) that first optimizes the muliti-objective programs to select stocks sector-wise on the basis of following four fixed FRs: 

 \begin{enumerate}
     \item Return on Assets (ROA) from the Profitability Ratio  
     \item Current Ratio (Liq) from the Liquidity Ratio 
     \item Debt-asset (Debt) Ratio from the Solvency Ratio  
     \item P/E Ratio (P/E) from the Valuation Ratio 
 \end{enumerate}

The selection of above ratios in each category is due to their popular usage in financial analysis. The optimal stocks found from the first step are then pooled together to generate a final optimal portfolio combining their performance in terms of the mean return and FRs. 
 \\

 \begin{center}
     \textbf{Step 1: Solving a multi-objective program for each sector}
 \end{center}
 In Step 1, $s$ optimization models are solved, separately for $s$ sectors, dominating the benchmark portfolio in SSD rule in constraints while keeping FRs in the objective functions. The models are multi-objective programs, named as $(SP1)_r; \ r=1,\ldots,s,$ defined as follows:
\\

 \begin{center}
 	$(SP1)_r$ \quad Max \ $\sum\limits_{i=1}^{n_r} (ROA)_{i}^{r}z_{i}^{r}$, \ Max	\ $\sum\limits_{i=1}^{n_r} (Liq)_{i}^{r}z_{i}^{r}$, \  Min \ $\sum\limits_{i=1}^{n_r} (Debt)_{i}^{r}z_{i}^{r}$, \ Min \ $\sum\limits_{i=1}^{n_r} (P/E)_{i}^{r}z_{i}^{r}$\\
 	 subject to \qquad \qquad \qquad \qquad \qquad \qquad \qquad \qquad \qquad \quad \qquad \quad \\
 	 \qquad \qquad $E(\eta^{r}-R_{z^r})^{+} \leq E(\eta^{r}-Y^{r})^{+}; \quad \eta^{r}=y_1^{r}, \ldots, y_T^{r}$\\
 	 \quad $0 \leq z_{i}^{r} \leq 0.3; \quad i=1, \ldots, n_r$ \\
 	 $\sum\limits_{i=1}^{n_r} z_i^r=1,$ \qquad \qquad 
 \end{center}
where $y_1^{r},\ldots, y_{T}^r$ are the $T$ realizations of $Y^{r},$ and 
$(ROA)_{i}^{r}, (Liq)_{i}^{r}, (Debt)_{i}^{r}$, and $(P/E)_{i}^{r}$ are respectively the profitability, liquidity, solvency, and valuation ratios of the $i$th asset from the $r$th sector. Constraints $z_{i}^{r} \leq 0.3$ impose a 30\% upper bound on the investment in each asset in every sector. Note that $\sum\limits_{r=1}^{s}n_{r}=N$, where N is the total number of assets. 

The $s$ multi-objective programs $(SP1)_r; \ r= 1,\ldots,s$ are converted into respective single objective programs $(WSP1)_r; \ r= 1,\ldots,s$ using the weighted-sum approach, given as:

\begin{align}
(WSP1)_r\hspace{0.2cm} & \text{Max}
\begin{aligned}[t]
\quad w^{r}_1\displaystyle \sum_{i=1}^{n_r}(ROA)_i^rz^r_i - w^{r}_2\displaystyle \sum_{i=1}^{n_r}(Debt)_i^rz^r_i+ w^{r}_3\displaystyle \sum_{i=1}^{n_r}(Liq)_i^rz^r_i - w^{r}_4\displaystyle \sum_{i=1}^{n_r}(P/E)_i^rz^r_i \\
\end{aligned}\notag\\
& \text{subject to} \notag \\
&E(\eta^r - R_{z^r})^+ \leq E(\eta^r - Y^r)^+;  \;\; \eta^r \, = \, y^r_1, \ldots,y^r_T\notag\\
&0 \; \leq z^r_i \leq 0.3; \;\; i=1,\ldots,n_r\notag\\
&\displaystyle\sum_{i=1}^{n_r}z^r_i = 1. \notag
\end{align}

The weights $w^{r}_j \; (j=1,2,3,4)$ in the programs $(WSP1)_r;\; r= 1,\ldots,s$ are obtained by analysing the effect of the individual FR on the Sharpe ratio, higher the effect of particular FR on Sharpe ratio, more weight (or importance) is given to that FR.

\begin{center}
    \textbf{Step 2: Combining stocks together}
 \end{center}
 
Step 2 of the SPO model combines optimal portfolio weights $(z^{r})^{*}; r= 1,\ldots,s$ derived from $s$ programs $(WSP1)_r; \; (r=1, \ldots, s)$ to generate final optimal portfolio.
Let $m_r=\{ i: \ (z_{i}^r)^{*} >0, 1 \leq i \leq n_{r} \}$ be the set comprising of filtered stocks from step 1, then the final optimal portfolio is obtained by solving the following optimization problem:

 \begin{center}
 	$(SP2)$ \quad Max \ $\sum\limits_{r=1}^{s}\sum\limits_{j \in m_r} \left( \left( \sum\limits_{t=1}^{T}r_{jt}p_{t} \right) + (z_{j}^{r})^{*} \right) z_{j}^{r}$ \\
 	subject to \qquad \qquad \qquad \quad \quad \quad \\
 	\qquad \qquad \qquad \quad $E(\eta-R_{z})^{+} \leq E(\eta-Y)^{+}; \ \eta=y_1, \ldots, y_T$\\
 	\qquad \qquad \qquad $0 \leq z_{j}^{r} \leq 0.3; \ j \in m_r, \ r=1, \ldots, s$\\
 	$\sum\limits_{r=1}^{s}\sum\limits_{j \in m_r} z_j^r=1,$
 \end{center}
where $y_1, y_2, \ldots , y_T$ are the $T$ discrete point realizations of benchmark portfolio $Y$. Step 2 ensures that the stocks having higher weights from Step 1 get higher allocation in the final optimal portfolio while dominating the benchmark portfolio $Y$ in SSD criteria when combined together.  

The SPO model incorporates the sector-wise use of FRs in the portfolio construction and has several advantages over the classical SSD model. However, it ignores the fact that the ratios that are found to be critical for one sector may not be so for another sector.  This is what formed the basis and primarily led to the following proposed strategy.

\section{Proposed Strategy: An extension to SPO}
\label{sec:4}

\subsection{Finding the Relevant ratios for each sector using PCA}
\label{sec:4.1}
We aim to improve the SPO model by carefully selecting the critical FRs in its Step 1 for each sector. To meet the objective, we propose to use Principal Component Analysis (PCA), a non-parametric statistical tool to select critical FRs separately for each sector.

PCA is the buttress of new age data analysis, devised by \citeauthor{pearson1901} \citeyearpar{pearson1901} as an analogue of principal axis theorem in mechanics and later developed and termed by \citeauthor{hotelling1933} \citeyearpar{hotelling1933}. PCA has been profusely used in several disciplines, from biomedical sciences to computer vision. The central idea of PCA is to compress a large data set of inter-related variables into a smaller set of variables while retaining maximum variation as in the original variables. PCA results into a new set of factors, called principal components which are uncorrelated and the first few components cumulatively explain most of the  variation (maximum information) as in the original variables. PCA on financial ratios was first applied by \citeauthor{pinches73} \citeyearpar{pinches73} with an objective to reduce the number of ratios used for predicting market crashes, bond ratings, and company failures.

PCA orthogonally transforms a data set of $k$ random variables, denoted as the random vector $X=\begin{bmatrix} X_1 & X_2 & \cdots & X_k \end{bmatrix}$ into a smaller set of linearly uncorrelated variables, called principal components. In this paper, the data set of $k$ variables corresponds to the $k$ financial ratios. If the $i$th ratio is denoted by $X_i$ with $m$ independent observations, then $X=\begin{bmatrix} X_1, & X_2, & \cdots & X_k \end{bmatrix}$ becomes a matrix of order $m \times k$, $X=[X_{ij}]$.

\textit{Steps for determining the Principal Components:}
\begin{enumerate}
    \item \textbf{Scaling of Data:} The mean row of the data matrix $X$ as $\overline{x}=\begin{bmatrix} \overline{x}_1 & \cdots &\overline{x}_k\end{bmatrix}$ of order $1 \times k$, where $\overline{x}_{j}=\dfrac{1}{m} \sum\limits_{i=1}^{m}X_{ij}\; \forall \; j =1,\ldots, k.$
    The mean matrix is then computed as $$\overline{X}=\mathbf{1}^{T} \overline{x} ,$$ where $\mathbf{1}$ denotes a row vector of ones of order $m.$ Subtracting $\overline{X}$ from the data matrix $X$ gives the scaled data matrix, $$Y=X-\overline{X}$$

    \item \textbf{Calculating the Covariance Matrix:} The covariance matrix is a $k \times k$ matrix and is given by $C=\dfrac{1}{m-1}Y^TY$. 
  
    \item \textbf{Computing the principal components:} To determine the principal components (PCs), we compute the eigenvalues and eigenvectors of the covariance matrix $C.$ Arranging the eigenvectors in descending order of their corresponding eigenvalues gives the principal components (PCs). The eigenvalues depicts the variance accounted by the corresponding PC (eigenvector).  
    
    \item \textbf{Deriving the loadings matrix:} After computing the principal components as eigenvectors, we compute the loadings matrix, comprising of component loadings (or simply loadings) defined as 
    \begin{center}
        Component Loadings$=\sqrt{\text{eigenvalue}} \times $ eigenvector
    \end{center}
    The component loadings denote the covariances (or correlations) between the original variables (here, ratios) and the principal components. 

\end{enumerate}

Sum of the squared values of the loadings gives the communality of the variable (ratio) with respect to the component solution, which denotes the variance in the original variable accounted for by the component solution. The extraction of the dominant ratios from PCA can be done in two ways, one on the basis of component loadings and two, on the basis of their communalties as discussed in detail in following section.

\begin{table}[h!]
\begin{center}
\footnotesize
    \caption{Financial Ratios used in the study }
    \begin{tabular*}{\textwidth}{@{\extracolsep{\fill}}lll@{\extracolsep{\fill}}}
\toprule
 Ratio Label   &  Ratio Name & Description \\
     \hline\noalign{\smallskip}
    A. Profitability Ratios$^{1}$ \\
    \hline\noalign{\smallskip}
     NPM & Net Profit Margin &  Net profit/Total revenue\\
     ROA & Return on Assets  &  Net Income/Total assets \\
     CPTI & Cash Profit Ratio &  Cash Profit/Total income \\
     ROE & Return on Equity  &  Net income/Average Shareholders' equity\\
    \toprule
    B. Liquidity Ratios$^{1}$ \\
    \hline\noalign{\smallskip}
     QR & Quick Ratio & Liquid Assets/Current Liabilities\\
     CR & Current Ratio & Current Assets/Current Liabilities\\
     CCL & Cash Ratio & Cash \& Cash equivalents/Current Liabilities \\
     \toprule
    C. Solvency Ratios$^{2}$  \\
    \hline\noalign{\smallskip}
     DER & Debt-Equity Ratio & Total debt/Shareholders' equity\\
     DAR & Debt-Asset Ratio & Total debt/Total assets\\
    \toprule
    D. Valuation Ratios$^{2}$   \\
    \hline\noalign{\smallskip}
     PER & P/E Ratio & Share price/Earnings per share (EPS)\\
     PBR & P/B Ratio & Market price per share/Book value per share
    \\
    \hline 
    \end{tabular*}
    \label{tab:Ratios}
\begin{tablenotes}%
\item[$^{1}$] larger value is preferable
\item[$^{2}$] smaller value is preferable
\end{tablenotes}
\end{center}
\end{table}

\subsection{Proposed strategies}
\label{sec:4.2}

For empirical purpose in the present study, PCA is applied to a set of $11$ financial ratios separately for each sector over a period of $10$ years from April 2004 to March 2014. The details of these 11 ratios is given in Table \ref{tab:Ratios}. We propose to fetch four components that explain atleast $80\%$ of the total ratio variances. These components together are referred to as the component solution. Though there is no thumb rule for extracting the significant original variables from the PCA solution, but several authors (for example, see \citeauthor{benchinyap} \citeyear{benchinyap}) have considered the loads\footnote{Loads gives the information of the amount of variance of the original variable(ratio), the principal component accounts for} from the loadings matrix as a measure of extraction.  

We adopt the following two methods for extracting the dominant ratios from PCs, explained as follows: 

\begin{itemize}
	\item[(A)] \textit{On the basis of component loadings}\\
	From each principal component, the financial ratio with the highest value (absolute) of component loading is extracted. This gives us the dominant ratios for each sector. These ratios may come from any of the category of the financial ratios, PR, LR, SR, and VR. However, we impose a bound on the extraction of maximum of two ratios from each category. 
	
	\item[(B)] \textit{On the basis of communalities}\\
	In this extraction method, we fetch the financial ratio having the maximum communality from each category, viz., PR, LR, SR and VR. In this way, we have four critical ratios that represent all four categories for each sector.  
\end{itemize}

After the selection of ratios for each sector, the two-step SPO model is employed to obtain final optimal portfolios. The portfolio selection strategy using extraction method (A) and extraction method (B), respectively are labelled as PCA-SPO(A) and PCA-SPO(B). The purpose of adopting the two extraction methods is to understand the importance and impact of inclusion of critical ratios representing all category of ratios on portfolio performance.

To examine the relevance of applying PCA sector-wise, we also apply PCA on the data when all assets are taken together to extract four dominant ratios. Four components that explain atleast $80\%$ of the total ratio variances are extracted. We then apply the nominal SSD model maximizing the weighted sum of the return function and the functions corresponding to the extracted FRs to obtain the optimal portfolio. We named this model as F-SSD. Let us denote the extracted ratios from PCA for the F-SSD model as $R_1, R_2, R_3$ and $R_4,$ then for $0\leq \alpha \leq 1,$ the F-SSD model is given as follows: 
\\
\begin{center}
	\textbf{(F-SSD)}  \quad    Max \ $\alpha (FR)_{z}+(1-\alpha) E(R_z)$\\
	\ subject to \qquad \qquad \qquad \ \\
	\qquad  \qquad \quad  \qquad \qquad \quad $E(\eta-R_{z})^{+} \leq E(\eta-Y)^{+}; \ \ \eta=y_1, \ldots, y_T$\\
	 $z \in Z,$
\end{center}
where
$$(FR)_z=\pm w_1\sum\limits_{j=1}^{N}(R_1)_jz_{j} \pm w_2\sum\limits_{j=1}^{N}(R_2)_jz_{j} \pm w_3\sum\limits_{j=1}^{N}(R_3)_jz_{j} \pm w_4\sum\limits_{j=1}^{N}(R_4)_jz_{j}.$$
The weights $w_1, w_2, w_3,$ and $w_4$ are taken as the proportion of variance the components (PCs) account for, where the sign (``$+$" for maximization and ``$-$" for minimization) indicates the effect of that ratio on the return. Here, $Z=\lbrace z=(z_1, z_2 \ldots, z_N): \ 0 \leq z_i \leq 0.3, \ \sum\limits_{i=1}^{N} z_i=1 \rbrace$
is the set of all admissible portfolios, imposing upper bound on each stock, with no short selling allowed and the budget constraint. 

For the current framework, we set $\alpha=0.5$ giving equal importance to returns and FRs. The F-SSD model incorporates FRs in the SSD framework when all stocks are considered together and does not focus on sector-wise bifurcation. Along with 
the F-SSD model, we also solve nominal SSD model which can be easily obtained from the F-SSD model for $\alpha=0$. The nominal SSD model maximizes only the return function of the portfolio and thus the assets are not chosen on the basis of their fundamental performance. In addition to SSD, the traditional Markowitz mean variance (Mean-var) and minimum variance (Min-var) models are also solved for the comparative analysis. Both the models minimize variance of a portfolio return in the objective function subject to the feasible region $z \in Z$ while the Mean-var model assumes an additional constraint on the expected return as $E(R_{z}) \geq \mu$, where $\mu$ is the expected return of the benchmark index.

\section{Empirical Analysis}
\label{sec:5}

\subsection{Sample data and sample period}
\label{sec:5.1}

We consider the two stock markets S\&P BSE 500 (India) and S\&P 500 (USA) as our sample data wherein we study following six sectors in the present empirical analysis.

\begin{enumerate}
    \item[(A)] Cyclic/Sensitive Sectors \qquad \qquad \qquad \qquad (B) Non-cyclic Sectors
    \begin{itemize}
        \item  [1] Energy \qquad \qquad \qquad \qquad \qquad \qquad  \qquad \qquad  1 Consumer Staples or FMCG 
        \item[2] Finance \qquad \qquad \qquad \qquad \qquad \ \ \quad \quad \quad \qquad  \ 2 HealthCare 
        \item[3] Information Technology (IT) \quad \quad \quad \ \quad \quad \quad \quad 3 Consumer Durables (CD) 
    \end{itemize}
\end{enumerate}

These sectors together represent approximately $70\%$ and $67\%$ of the whole market of S\&P BSE 500 and S\&P 500, respectively. 

Bombay Stock Exchange (BSE) is the oldest stock exchange in Asia and the $10$th largest in the world by market capitalization as on March, 2020. On February 19, 2013, BSE partnered with S\&P DOW JONES INDICES and got renamed as S\&P BSE ``*" (* indicates various defined sets like SENSEX, 100, etc). The S\&P BSE 500 comprises of 500 companies covering all major industry sectors of the Indian economy and depicts nearly 93\% of BSE's total market capitalization. On the other hand, the S\&P 500 is a stock market comprising of the 500 large capitalization companies listed on the stock exchanges in the United States, awning over the major industry sectors. As of September 2021, the top nine companies of the index including Apple, Microsoft, Amazon, Tesla and Berkshire Hathway among others accounted for about 28.1\% of the total market capitalization of the index.

We collect the following sample data for stocks listed under the above mentioned 6 sectors of S\&P BSE 500 (India) from CMIE Prowess IQ Database (https://prowessiq.cmie.com/) and of S\&P 500 (U.S.) from Bloomberg Database Management System. 
\begin{enumerate}
    \item The weekly adjusted closing price from April 2014-March 2020.
    \item Quarterly data of all the ratios listed in Table \ref{tab:Ratios} from April 2004-March 2019, giving a total of 40 data points for each asset and each ratio. Thus, for sector $r$ with number of assets $n_r,$ the sample data is a matrix with $(n_r \times 40)$ rows and $11$ columns (for ratios) for proposed strategies and a matrix with $115 \times 40$ rows and 11 columns in case of F-SSD.
\end{enumerate}

The weekly return for $j$-th stock in $t$-th week is calculated as $r_{jt}=\frac{p_{jt}^{c}-p_{jt-1}^{c}}{p_{jt-1}^{c}};$ $j=1,\ldots, N, \ t=1,\ldots, T,$  where $p_{jt}^{c}$ and $p_{jt-1}^{c}$ are respectively, the closing prices in the $t$-th and $(t-1)$-th week. We use R software with SYMPHONY solver on Windows 64 bits Intel(R) Core(TM) i3-6006U CPU @2.0 GHz processor to solve all optimization models, except Mean-var and Min-var for which we use the CVXR package in R.  

Stocks whose price data is missing for more than 2 quarters and/or whose ratios data is missing for more than 4 quarters are deleted from the study. The total number of assets henceforth in the present study are left to $115$ for the Indian market and $179$ for the U.S. market, whose sector-wise composition is listed in Table \ref{tab:sectorcomp}.

\begin{table}[h!]
    \centering
     \caption{Sector-wise composition of the S\&P BSE 500 and S\&P 500 data sets}
      \begin{tabular*}{\textwidth}{@{\extracolsep{\fill}}c|ccccccc@{\extracolsep{\fill}}}
     \toprule
        Market & Energy & Finance & IT & FMCG & Healthcare & CD & Total \\
         \hline\noalign{\smallskip}
         S\&P 500 (U.S.) & 15 & 10 & 41 & 26 & 49 & 38 & 179\\
         S\&P BSE 500 (India) & 14 & 32 & 12 & 23 & 27 & 7 & 115\\
         \hline 
    \end{tabular*}
    \label{tab:sectorcomp}
\end{table}

To analyze the financial benefits of the proposed strategies,  PCA-SPO(A), PCA-SPO(B) vis-vis the models, SPO, SSD, F-SSD, Mean-var, Min-var, we perform the analysis in two ways. Firstly, by using a rolling window scheme and second by analyzing their performance over different phases (price movement) of the market trends, viz., neutral (up-down-up movements), bullish (upwards movement) and bearish (downwards movement) for both the data sets. 

The performance of the optimal portfolios obtained from all the models are compared on the basis of following performance measures:

\begin{itemize}
    \item \textbf{Mean Return}: It is the average of the $M$ out-of-sample portfolio returns $R_{zt}$; where $M$ denotes the total number of out-of-sample windows. It is given as: 
   
    $$MR=\dfrac{1}{M} \sum\limits_{t=1}^{M} R_{zt}=\dfrac{1}{M} \sum\limits_{t=1}^{M} \sum\limits_{j=1}^{N} r_{jt}z_{j}$$
    Higher values of Mean Return are preferable.
    
    \item \textbf{Downside Deviation}: It is defined as
    $$DD=\dfrac{1}{\sqrt{M}} \ \sqrt{\sum\limits_{t=1}^{M} \text{min}(R_{zt},0)^2},$$ where $M$ is the total number of out-of-sample windows. Its lower values are desirable.
    
    \item \textbf{VaR$_\alpha$ and CVaR$_{\alpha}$}: It denotes the sample Value at Risk (VaR) and sample Conditional VaR values for $\alpha\%$ level of confidence. Arranging the out-of-sample returns from portfolio $z$ in ascending order as $R_{z1}, R_{z2}, \ldots, R_{zM};$ where $M$ is the total number of out-of-sample windows, VaR and CVaR at $\alpha\%$ level of confidence are given as 
    \begin{align*}
        \text{VaR}_{\alpha}(R_{z}) &=R_{zk}\\
        \text{CVaR}_{\alpha}(R_{z}) &=\frac{1}{M(1-\alpha)} \sum\limits_{i=1}^{k} R_{zi}
    \end{align*}
    where $k=\lfloor{M(1-\alpha)}\rfloor{}+1$. Here, $\lfloor{\cdot}\rfloor{}$ denotes the greatest integer function or the floor function. 
    
    \item \textbf{Sharpe Ratio}: It is the ratio of expected excess return over risk-free return $r_{f}$ to the standard deviation of portfolio returns $\sigma_{R_{z}}$. 
    $$\text{Sharpe}=\dfrac{E(R_{z})-r_{f}}{\sigma_{R_{z}}},$$ 
    where $E(R_{z})-r_{f}>0$. Higher values are worthwhile.
    
    \item \textbf{Sortino Ratio}: It is the ratio of expected excess return over risk-free return to the downside deviation $DD$.  $$\text{Sortino}= \dfrac{E(R_{z})-r_{f}}{{DD}},$$ where $E(R_{z})-r_{f}>0$. 
    Higher values are preferred.
    
    \item \textbf{Rachev Ratio}: It is the ratio of expected tail returns in the best $(1-\alpha)\%$ cases to the expected tail losses in the worst $(1-\alpha)\%$ cases. It is given as  $$\text{Rachev}_{\alpha}=\dfrac{\text{CVaR}_{\alpha}(R_{z})}{\text{CVaR}_{\alpha}(-R_{z})}.$$ Larger values are desirable.
    
    \item \textbf{STARR Ratio}: STARR ratio is the ratio of expected excess return over risk-free return to its CVaR$_{\alpha}$ value. $$\text{STARR}_{\alpha}=\dfrac{E(R_{z})-r_{f}}{\text{CVaR}_{\alpha}(R_{z})},$$ where $E(R_{z})-r_{f}>0$. Larger values are preferable. ratios. 
    
    We report the values of the VaR, CVaR, Rachev and the STARR ratios for $95\%$ and $97\%$ levels of confidence.
    
\end{itemize}

\subsection{In-sample outcomes}
\label{sec:5.2}
Recall that four components that explain atleast $80\%$ of the total ratio variances are extracted for the proposed strategies, as well as for the F-SSD model. Moreover, ratios having the highest component load are selected from each principal component in case of PCA-SPO(A) and F-SSD whereas ratios having maximum communality from each category are selected for PCA-SPO(B). 

Table \ref{tab:secloadind} illustrates the values of component loading for the CD and Energy sectors of the S\&P BSE 500 data set, while Table \ref{tab:secloadus} illustrates the component loadings for the FMCG and IT sectors of the S\&P 500 data set. The four extracted components (PC1-PC4) are referred as the component solution. We can note that four principal components that are extracted for the sectors CD and Energy of the S\&P BSE 500 market together explain respectively, about $94.01\%$ and $90.23\%$ while those for the sectors FMCG and IT of the S\&P 500 market together explain 85.53\% and 90.75\% of the total ratio variances respectively. The last column denotes the communalities (denoted as Comm. in Tables \ref{tab:secloadind} and \ref{tab:secloadus}) of the ratios accounted by the component solution. The sign of the loading is indicative of the direction of relationship and is thus ignored while fetching the ratios from the components.

From Table \ref{tab:secloadind}, we observe that the ratios CR, ROA, PER, and CCL are found to be the critical ratios for the CD sector (highlighted in bold) having respective loading as 0.9775, 0.9158, $-0.6654$, and $-0.6576$ for PCA-SPO(A) strategy. And for the strategy PCA-SPO(B), the critical ratios for the CD sector becomes CR, ROE, DAR and PBR (bold \& italicised) having respective communalities as 0.9759, 0.9839, 0.9794, and 0.9173. Similar readings are for the Energy sector from Table \ref{tab:secloadind} and for the FMCG and IT sectors from Table \ref{tab:secloadus}.

\begin{table}[!ht]
 \centering
    \caption{Component loadings matrices for CD and Energy Sector for S\&P BSE-500 dataset}
    \label{tab:secloadind}
     \footnotesize
     \begin{tabular*}{\textwidth}{@{\extracolsep{\fill}}l|ccccc|ccccc@{\extracolsep{\fill}}}
\hline\noalign{\smallskip}
\multirow{2}{*}{Ratios} & \multicolumn{5}{c|}{Consumer Durables} & \multicolumn{5}{c}{Energy} \\
 & PC1 & PC2 & PC3 & PC4 & Comm. & PC1 & PC2 & PC3 & PC4 & Comm.\\
\noalign{\smallskip}\hline\noalign{\smallskip}
\multicolumn{11}{l}{LR}  \\
\hline\noalign{\smallskip}
QR & 0.8885 & $-0.2441$ & $-0.1309$ & $-0.1369$ & 0.8849 & $-0.9267$ & 0.0443 & $-0.3293$ & $-0.0159$ & 0.9695\\
CR & \bf{0.9775} & $-0.0172$ & $-0.1393$ & $-0.0261$ & \bf{\emph{0.9759}} & $-0.7734$ & 0.3095 & $-0.4831$ & $-0.0592$ & 0.9309\\
CCL & 0.5454 & 0.3073 &  0.3860 & $\mathbf{-0.6576}$ & 0.9734 & $\mathbf{-0.9271}$ & 0.1509 & $-0.3199$ & 0.0512 & \bf{\emph{0.9873}}\\
\noalign{\smallskip}\hline\noalign{\smallskip}
\multicolumn{11}{l}{PR}\\
\hline\noalign{\smallskip}
NPM & 0.4540 & 0.8279 & -0.0078 & 0.2826 & 0.9716 & $-0.7117$ & 0.3077 & 0.1476 & $-0.4258$ & 0.8043\\
ROA & 0.2508 & \bf{0.9159} & 0.0049 & 0.2801 & 0.9802 & $-0.8114$ & 0.2349 & 0.4572 & $-0.1343$ & \bf{\emph{0.9406}}\\
CPTI & $-0.2493$ & 0.8533 & 0.2938 & $-0.0169$ & 0.8769 & $-0.8958$ & $-0.0007$ & 0.0191 & 0.2529 & 0.8669\\
ROE & $-0.5865$ & 0.7787 & 0.1791 & $-0.0373$ & \bf{\emph{0.9839}} & $-0.1725$ & 0.4629 & \bf{0.7953} & $-0.0688$ & 0.8813\\
\noalign{\smallskip}\hline\noalign{\smallskip}
\multicolumn{11}{l}{SR}\\
\hline\noalign{\smallskip}
DER & $-0.7159$ & $-0.5552$ & 0.2339 & 0.0966 & 0.8848 & $-0.2436$ & 0.6389 & 0.2140 & \bf{0.6076} & \bf{\emph{0.8826}}\\
DAR & $-0.9343$ & $-0.2002$ & 0.2488 & $-0.0675$ & \bf{\emph{0.9794}} & 0.5648 & 0.6838 & $-0.1598$ & 0.1275 & 0.8284\\
\noalign{\smallskip}\hline\noalign{\smallskip}
\multicolumn{11}{l}{VR}\\
\hline\noalign{\smallskip}
PER & $-0.6339$ & 0.1966 & $\mathbf{-0.6654}$ & $-0.1718$ & 0.9127 & $-0.3858$ & $-0.8310$ & 0.0843 & 0.2609 & 0.9147\\
PBR & $-0.5358$ & 0.6158 & $-0.3746$ & $-0.3327$ & \bf{\emph{0.9173}} & $-0.4787$ & $\mathbf{-0.7479}$ & 0.3557 & 0.0705 & \bf{\emph{0.9200}}\\
\noalign{\smallskip}\hline\noalign{\smallskip}
CV & 0.4356 & 0.7792 & 0.8705 & 0.9401 & &  0.4609 & 0.6982 & 0.8357 & 0.9023\\
\noalign{\smallskip}\hline
\end{tabular*}
\end{table}

\begin{table}[htp!]
\centering
    \caption{Component loadings matrices for FMCG and IT Sectors for S\&P 500 dataset}
    \footnotesize
    \label{tab:secloadus}
     \begin{tabular*}{\textwidth}{@{\extracolsep{\fill}}l|ccccc|ccccc@{\extracolsep{\fill}}}
\hline\noalign{\smallskip}
\multirow{2}{*}{Ratios} & \multicolumn{5}{c|}{FMCG} & \multicolumn{5}{c}{Information Technology} \\
 & PC1 & PC2 & PC3 & PC4 & Comm. & PC1 & PC2 & PC3 & PC4 & Comm.\\
\noalign{\smallskip}\hline\noalign{\smallskip}
\multicolumn{11}{l}{LR}  \\
\hline\noalign{\smallskip}
QR & 0.8773 & $-0.3488$ & 0.2389 & $-0.0145$ & \bf{\emph{0.9486}} & $-0.8693$ & $-0.0511$ & 0.0423 & $\mathbf{-0.4727}$ & 0.9835\\
CR & 0.8740 & $-0.2769$ & 0.3235 & $-0.0005$ & 0.9452 & $\mathbf{-0.8846}$ & $-0.0887$ & 0.0953 & $-0.4249$ & 0.9800\\
CCL & \bf{0.8919} & $-0.2811$ &  0.2027 & $-0.0370$ & 0.9169 & $-0.8783$ & $-0.0476$ & 0.0896 & $-0.4562$ & \bf{\emph{0.9898}}\\
\noalign{\smallskip}\hline\noalign{\smallskip}
\multicolumn{11}{l}{PR}\\
\hline\noalign{\smallskip}
NPM & 0.2058 & $-0.5475$ & 0.5283 & $-0.0408$ & 0.6229 & $-0.7575$ & 0.0245 & 0.1520 & 0.3992 & 0.7569\\
ROA & $-0.6120$ & $-0.6399$ & 0.2576 & $-0.3053$ & 0.9436 & $-0.8741$ & 0.0469 & 0.1337 & 0.2826 & \bf{\emph{0.8639}}\\
CPTI & 0.0684 & 0.1720 & $-0.3624$ & $\mathbf{-0.8324}$ & 0.8585 & 0.5876 & $-0.1115$ & 0.2518 & $-0.4706$ & 0.6426\\
ROE & $-0.6580$ & $-0.6021$ & 0.2256 & $-0.3152$ & \bf{\emph{0.9457}} & $-0.7924$ & 0.2379 & 0.1444 & 0.3921 & 0.8591\\
\noalign{\smallskip}\hline\noalign{\smallskip}
\multicolumn{11}{l}{SR}\\
\hline\noalign{\smallskip}
DER & $-0.6509$ & 0.4137 & 0.5515 & 0.0653 & 0.9032 & 0.2174 & \bf{0.9110} & 0.3194 & $-0.1065$ & \bf{\emph{0.9905}}\\
DAR & $-0.6567$ & 0.3742 & \bf{0.5794} & 0.0547 & \bf{\emph{0.9099}} & 0.1854 & 0.9099 & 0.3392 & $-0.1083$ & 0.9891\\
\noalign{\smallskip}\hline\noalign{\smallskip}
\multicolumn{11}{l}{VR}\\
\hline\noalign{\smallskip}
PER & $-0.3401$ & $-0.6079$ & $-0.3974$ & 0.3841 & 0.7907 & 0.0503 & 0.2616 & $\mathbf{-0.9440}$ & $-0.1440$ & \bf{\emph{0.9828}}\\
PBR & $-0.5738$ & $\mathbf{-0.7321}$ & $-0.2672$ & 0.1241 & \bf{\emph{0.9531}} & $-0.4922$ & 0.5446 & $-0.6364$ & 0.0007 & 0.9439\\
\hline\noalign{\smallskip}
CV & 0.4076 & 0.6435 & 0.7889 & 0.8553 &  & 0.4497  & 0.6413 & 0.7920 &0.9075 & \\
\noalign{\smallskip}\hline\noalign{\smallskip}
\end{tabular*}
\end{table}

Tables \ref{tab:ratiosA} and \ref{tab:ratiosB} respectively, exhibit the ratios selected for both the proposed strategies PCA-SPO(A) and PCA-SPO(B) corresponding to all sectors for both the indices. The last columns of the Tables describe the cumulative variance (denoted as CV) accounted by the component solution for each sector. 

\begin{table}[htp!]
\centering
 \footnotesize
    \caption{Critical Ratios for each sector for the strategy PCA-SPO(A)}
    \label{tab:ratiosA}
    \begin{tabular*}{\textwidth}{@{\extracolsep{\fill}}lllllc|llllc@{\extracolsep{\fill}}}
\hline\noalign{\smallskip}
\multirow{2}{*}{Sectors} & \multicolumn{5}{c|}{S\&P BSE 500} & \multicolumn{5}{c}{S\&P 500}\\
 & R1 & R2 & R3 & R4 & CV & R1 & R2 & R3 & R4 & CV\\
\hline\noalign{\smallskip}
Energy & CCL & PER & ROE & DER & 90.23\% & ROA & DAR & PBR & CPTI & 87.96\% \\
Finance & ROE & CCL & PER & DAR & 93.99\% & CCL & DAR & PBR & CPTI & 84.23\%\\
FMCG & DAR & NPM & PER & DER & 94.69\% & CCL & PBR & DAR & CPTI & 88.53\%\\
CD & CR & ROA & PER & CCL & 94.01\% & CCL & PER & PBR & CPTI & 82.91\%\\
IT & CR & PBR & ROE & ROA & 90.81\% & CR & DER & PER & QR & 90.75\%\\
HC & CCL & DER & PBR & PER & 90.04\% & QR & PBR & NPM & CPTI & 92.41\%\\
\noalign{\smallskip}\hline
\end{tabular*}
\end{table}

\begin{table}[htp!]
\centering
 \footnotesize
    \caption{Critical Ratios for each sector for the strategy PCA-SPO(B)}
    \label{tab:ratiosB}
    \begin{tabular*}{\textwidth}{@{\extracolsep{\fill}}lllllc|llllc@{\extracolsep{\fill}}}
\hline\noalign{\smallskip}
\multirow{2}{*}{Sectors} & \multicolumn{5}{c|}{S\&P BSE 500} & \multicolumn{5}{c}{S\&P 500}\\
 & LR & PR & SR & VR & CV & LR & PR & SR & VR & CV\\
\noalign{\smallskip}\hline\noalign{\smallskip}
Energy & CCL & ROA & DER & PBR & 90.23\% & QR & ROA & DER & PER & 87.96\% \\
Finance & CCL & ROE & DER & PER & 93.99\% & CCL & ROE & DER & PER & 84.23\%\\
FMCG & CCL & ROE & DER & PBR & 94.69\% & QR & ROE & DAR & PBR & 88.53\%\\
CD & CR & ROE & DAR & PBR & 94.01\% & QR & ROE & DER & PER & 82.91\%\\
IT & CR & ROA & DER & PBR & 90.81\% & CCL & ROE & DER & PER & 90.75\%\\
HC & QR & NPM & DER & PER & 90.04\% & QR & CPTI & DER & PBR & 92.41\% \\
\noalign{\smallskip}\hline
\end{tabular*}
\end{table}

Table \ref{tab:loadall} displays the values of component loadings when PCA is applied on all assets at once. 
The dominant ratios for the Indian S\&P BSE 500 dataset are obtained as CR, DER, PBR and CPTI having respectively, 0.8561, 0.7535, $-0.7060$, and $-0.5011$ loads, while for that of the U.S. S\&P 500 are ROA, ROE, CCL and CPTI having respective loading as $-0.8670$, $-0.6899$, $0.5814$, and $-0.7228$. The last row represents the proportion of variance each principal component accounts for and are taken as the weights corresponding to each ratio in the F-SSD model. It is interesting to note that while for S\&P BSE 500, ratios from each category are picked up in the selection process, the profitability ratios dominate the selection for S\&P 500 index. 

\begin{table}[htp!]
    \centering
    \caption{Factor loadings matrices for ALL assets for S\&P BSE-500 and S\&P 500 datasets}
    \footnotesize
    \label{tab:loadall}
  \begin{tabular*}{\textwidth}{@{\extracolsep{\fill}}l|cccc |cccc@{\extracolsep{\fill}}}
\hline\noalign{\smallskip}
\multirow{2}{*}{Ratios} & \multicolumn{4}{c|}{S\&P BSE 500} & \multicolumn{4}{c}{S\&P 500 } \\
 & PC1 & PC2 & PC3 & PC4 & PC1 & PC2 & PC3 & PC4 \\
\noalign{\smallskip}\hline\noalign{\smallskip}
\multicolumn{9}{l}{LR}  \\
\hline\noalign{\smallskip}
QR & 0.8545 & 0.4003 & $-0.1891$ & 0.1594 & 0.8037 & $-0.0388$ & 0.5774 & $-0.0273$ \\
CR & \bf{0.8561} & 0.3577 & $-0.1747$ & 0.2099 & 0.7949 & $-0.1062$ & 0.5656 & $-0.0203$ \\
CCL & 0.8149 & 0.3868 & $-0.2745$ & 0.0765 & 0.7466 & 0.1750 & \bf{0.5814} & 0.0253 \\
\noalign{\smallskip}\hline\noalign{\smallskip}
\multicolumn{9}{l}{PR}\\
\hline\noalign{\smallskip}
NPM & 0.8301 & $-0.4167$ & $-0.0330$ & $-0.1462$ & 0.5826 & 0.3395 & $-0.2295$ & \bf{0.3387}\\
ROA & 0.6691 & $-0.6311$ & 0.1598 & 0.0912 & \bf{0.8670} & $-0.2939$ & $-0.1760$ & $-0.0604$ \\
CPTI & 0.5896 & 0.2340 & $-0.4255$ & $\mathbf{-0.5011}$ & $-0.3545$ & 0.4351 & 0.1452 & $-0.7228$\\
ROE & 0.6559 & $-0.5815$ & 0.1344 & $-0.0923$ & 0.5159 & $\mathbf{-0.6899}$ & $-0.2675$ & $-0.0646$\\
\noalign{\smallskip}\hline\noalign{\smallskip}
\multicolumn{9}{l}{SR} \\
\hline\noalign{\smallskip}
DER & $-0.3616$ & \bf{0.7535} & $-0.1806$ & $-0.2608$ & $-0.6486$ & $-0.6232$ & 0.2301 & $-0.0505$\\
DAR & $-0.4178$ & $-0.2654$ & $-0.6889$ & 0.1017 & $-0.7412$ & $-0.5188$ & 0.3391 & $-0.0349$\\
\noalign{\smallskip}\hline\noalign{\smallskip}
\multicolumn{9}{l}{VR}\\
\noalign{\smallskip}\hline\noalign{\smallskip}
PER & $-0.2069$ & $-0.5821$ & $-0.3829$ & $-0.4210$ & 0.6849 & 0.0731 & $-0.3842$ & $-0.3035$\\
PBR & $-0.2629$ & $-0.3390$ & $\mathbf{-0.7060}$ & 0.3833 & 0.7482 & $-0.3832$ & $-0.2782$  & $-0.2666$\\
\noalign{\smallskip}\hline\noalign{\smallskip}
Prop. of Variance & 0.4054 & 0.2268 & 0.1381 & 0.0697 & 0.4834 & 0.1564 & 0.1419 & 0.0739  \\
\hline
\end{tabular*}
\end{table}

We now analyze the financial benefits of proposed strategies PCA-SPO(A) and PCA-SPO(B) in comparison to five models, viz., SPO, F-SSD, the nominal SSD, Mean-variance (Mean-var) and Minimum variance (Min-var) on the basis of their out-of-sample findings over both the data sets.

\subsection{Out-of-sample analysis}
\label{sec: 5.3}

In this section, we carry out the out-of-sample analysis, conducted in two ways; firstly by using a rolling window scheme and second by using scenario based analysis on the different market price movement, viz., neutral (up-down-up movements), bullish (upwards movement) and bearish (downwards movement) for both the markets: S\&P BSE 500 (India) and S\&P 500 (US). 

\subsubsection{Out-of-sample analysis using rolling window scheme}

We follow a rolling window scheme of 4 weeks over the time horizon of 6 years from April 2014 to March 2020 with in-sample period consisting of 52 weeks and out-of-sample period of 3 months or 13 weeks. Sliding the in-sample period by 4 weeks, we get a total of 51 out-of-sample windows, or equivalently 663 $(51 \times 13)$ out-of-sample returns for both the data sets.

Table \ref{tab:osp3} records the performance of the portfolios derived from all the models using the rolling window scheme on the Indian S\&P BSE 500 index. Following observations are drawn from Table \ref{tab:osp3}:

\begin{enumerate}
    \item The portfolio from PCA-SPO(A) outperforms all other  portfolios in terms of mean return, Sharpe ratio, and the STARR ratios.
    
    \item The portfolio from PCA-SPO(B) outperforms all other portfolios in terms of all risk measures, the Sortino ratio and the Rachev ratios whereas it takes the second best position in terms of the Sharpe and STARR ratios after PCA-SPO(A). 
    
    \item The F-SSD model turned out to be the most risky as it suffers with the highest values of DD, CVaR$_{0.95}$, CVaR$_{0.97}$, VaR$_{0.95}$, and VaR$_{0.97}$ in comparison to other portfolios, which hints that analysing all stocks together on the basis of FRs may mislead an investor. Moreover, not using FRs at all in the selection process can also account for high risk as SSD model stands the second most risky model.     

   \item None of the proposed strategies, PCA-SPO(A) and PCA-SPO(B) generated worst results in any of the performance measures considered in the study. 
\end{enumerate}

\vspace{-5pt}

\begin{table}[htp!]
    \centering
    \caption{Out-of-sample performance of portfolios for S\&P BSE 500 (India) data set using rolling window strategy}
    \footnotesize
    \label{tab:osp3}
    \tabcolsep=0pt
  \begin{tabular*}{\textwidth}{@{\extracolsep{\fill}}lccccccc@{\extracolsep{\fill}}}
\noalign{\smallskip}\hline\noalign{\smallskip}
Portfolios & SSD & SPO & Min-Var & Mean-Var &  PCA-SPO(A) & PCA-SPO(B) & F-SSD  \\
\noalign{\smallskip}\hline\noalign{\smallskip}
Mean Return & 0.00384 & 0.00364 & 0.00201 & 0.00201 & 0.00232 & \bf{0.00471} & 0.00375 \\
Sharpe Ratio  & 0.29883 & 0.42849 & 0.18519 & 0.19169 & 0.13494 & \bf{0.51429} & 0.47407 \\
Sortino Ratio & 0.66021 & 0.96805 & 0.35762 & 0.36272  & 0.23782 & 1.59761 & \bf{1.60163}\\
DD  & 0.00389 & 0.00245 & 0.00208 & 0.00202 & 0.00442 & 0.00215  & \bf{0.00155} \\
VaR$_{0.95}$ & 0.01015 & 0.00639  & 0.00558 & 0.00558 & 0.01087  &  0.00535 & \bf{0.00373} \\
VaR$_{0.97}$ & 0.01113 & 0.00648 & 0.00629 & 0.00655 &  0.01211 & 0.00598 & \bf{0.00542} \\
CVaR$_{0.95}$  & 0.01104 & 0.00807 & 0.00676 & 0.00685 & 0.01250 & 0.00633 & \bf{0.00497} \\
CVaR$_{0.97}$ & 0.01149 & 0.00891 & 0.00735 & 0.00748 & 0.01380 &  0.00681 & \bf{0.00558} \\
Rachev$_{0.95}$  & 1.86926 & 1.80834 & 1.21052 & 1.13012 & 1.03467 & 3.02685 & \bf{3.24614} \\
Rachev$_{0.97}$  & 1.91427 & 1.72559 & 1.16076 & 1.04097 & 0.95072 & 2.92070 & \bf{3.22673} \\
STARR$_{0.95}$ & 0.23263 & 0.29389 & 0.10980 & 0.10728 & 0.08409 & \bf{0.54263} & 0.49950 \\
STARR$_{0.97}$ & 0.22352 & 0.26619 & 0.10102 & 0.09823 & 0.07617 & \bf{0.50438} & 0.44489 \\
\noalign{\smallskip}\hline
\end{tabular*}
\end{table}

Table \ref{tab:osp3us} records the performance of portfolios obtained from all the models on the USA S\&P 500 market having following observations:

\begin{enumerate}
    \item Portfolios from PCA-SPO(A) outperforms all portfolios in terms Sharpe, Sortino and STARR ratios. 
    
    \item Portfolios from PCA-SPO(B) outperforms all portfolios in terms of VaR values and takes the second best position in terms of Sharpe and Sortino ratios.
    
    \item Portfolios from the proposed strategies: PCA-SPO(A) and PCA-SPO(B) outperforms portfolios from SPO model in terms of mean return, Sharpe, Sortino, STARR ratios as well as risk measures: DD, VaR$_{0.95}$, VaR$_{0.97}$, CVaR$_{0.95}$ and CVaR$_{0.97}$ values.
    
    \item Portfolios from the F-SSD model generate highest mean return but suffers with the highest risks in terms of DD, VaR$_{0.95}$, VaR$_{0.97}$, CVaR$_{0.95}$ and CVaR$_{0.97}$ values.

    \item Here as well, none of the proposed strategies suffer with the worst outcome in any of the performance measures.

\end{enumerate}
  
\vspace{-5pt}

\begin{table}[htp!]
    \centering
    \caption{Out-of-sample performance of portfolios of S\&P 500 (U.S.) data set using rolling window strategy}
    \footnotesize
    \label{tab:osp3us}
    \tabcolsep=0pt
  \begin{tabular*}{\textwidth}{@{\extracolsep{\fill}}lccccccc@{\extracolsep{\fill}}}
\noalign{\smallskip}\hline\noalign{\smallskip}
Portfolios & SSD & SPO & Min-Var & Mean-Var &  PCA-SPO(A) & PCA-SPO(B) & F-SSD  \\
\noalign{\smallskip}\hline\noalign{\smallskip}
Mean Return & 0.00333 & 0.00215 & 0.00136 & 0.00150 & \textbf{0.00477} & 0.00277 & 0.00245 \\
Sharpe Ratio & 0.44419 & 0.41995 & 0.34369 & 0.38487 & 0.46824 & \textbf{0.54088} & 0.48840 \\
Sortino Ratio & 0.91133 & 0.72415 & 0.63283 & 0.72599 & 0.77402 & \textbf{1.08074} & 0.81523 \\
DD & 0.00365 & 0.00297 & 0.00215 & \textbf{0.00206} & 0.00577 & 0.00257  & 0.00300 \\
VaR$_{0.95}$ & 0.00869  & 0.00737 & 0.00527  & 0.00527 & 0.01630 &  0.00593 & \textbf{0.00353} \\
VaR$_{0.97}$ & 0.01040 & 0.00752 & 0.00709 & 0.00717 & 0.02003 & 0.00648 & \textbf{0.00508} \\
CVaR$_{0.95}$ & 0.01130 & 0.00939 & \textbf{0.00800} & 0.00783 & 0.02559 & 0.00892  & 0.00937 \\
CVaR$_{0.97}$ & 0.01260 & 0.01041 & \textbf{0.00989} & 0.00960 & 0.03201 & 0.01042 & 0.01229 \\
Rachev$_{0.95}$ & \textbf{1.53910} & 1.13193 & 1.39088 & 1.41761 & 0.81772 & 1.21299 & 1.06277 \\
Rachev$_{0.97}$ & \textbf{1.44630} & 1.03608 & 1.36301 & 1.40387 & 0.77525 & 1.08396 & 0.82853 \\
STARR$_{0.95}$ & 0.29466 & 0.22882 & 0.17034 & 0.19112 & 0.17455 & \textbf{0.31089} & 0.26118 \\
STARR$_{0.97}$ & 0.26415 & 0.20657 & 0.13777 & 0.15577 & 0.13958 & \textbf{0.26621} & 0.19914 \\
\noalign{\smallskip}\hline
\end{tabular*}
\end{table}

\begin{figure}[!ht]
  \includegraphics[height=6cm, width=\textwidth]{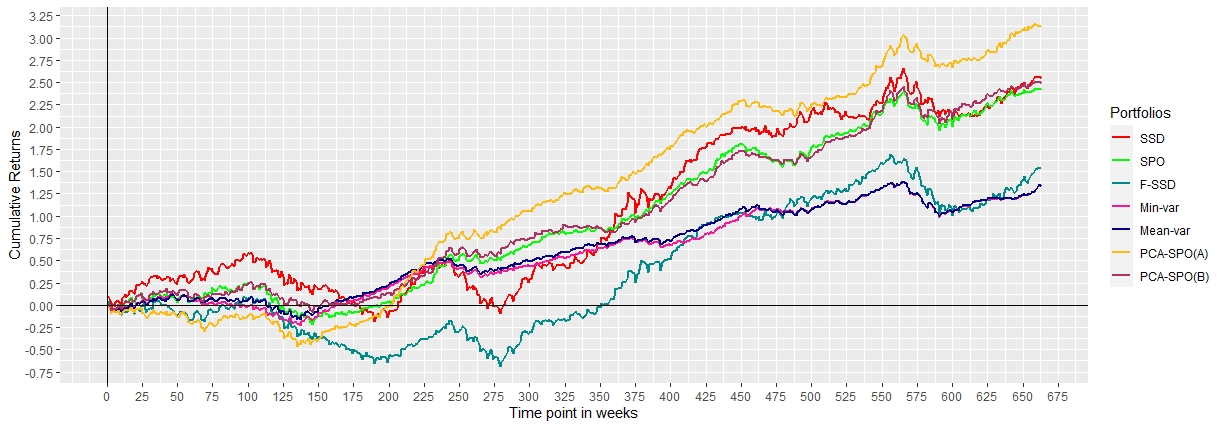}
  \caption{Cumulative returns of the portfolios from PCA-SPO(A), PCA-SPO(B), SPO, SSD, F-SSD, Min-var and Mean-var for S\&P BSE 500 Market (India) for 3 month out-of-sample period (rolling window scheme)}
  \label{fig:CR_IND}
\end{figure}
 
\begin{figure}[!ht]
  \includegraphics[height=6cm, width=\textwidth]{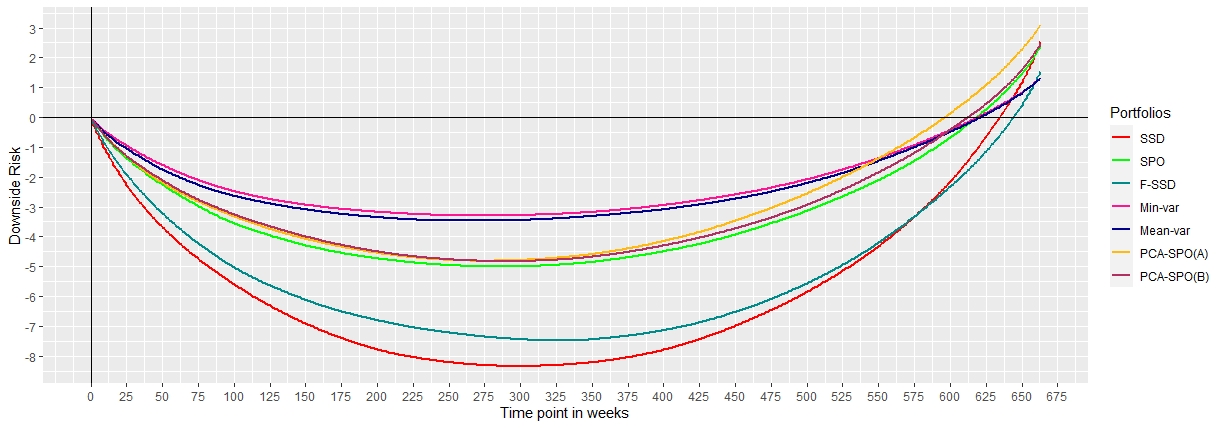}
  \caption{Downside Risk of the portfolios from PCA-SPO(A), PCA-SPO(B), SPO, SSD, F-SSD, Min-var and Mean-var for S\&P BSE 500 Market (India) for 3 month out-of-sample period (rolling window scheme)}
  \label{fig:DR_IND}
\end{figure}

\begin{figure}[!ht]
  \includegraphics[height=6cm, width=\textwidth]{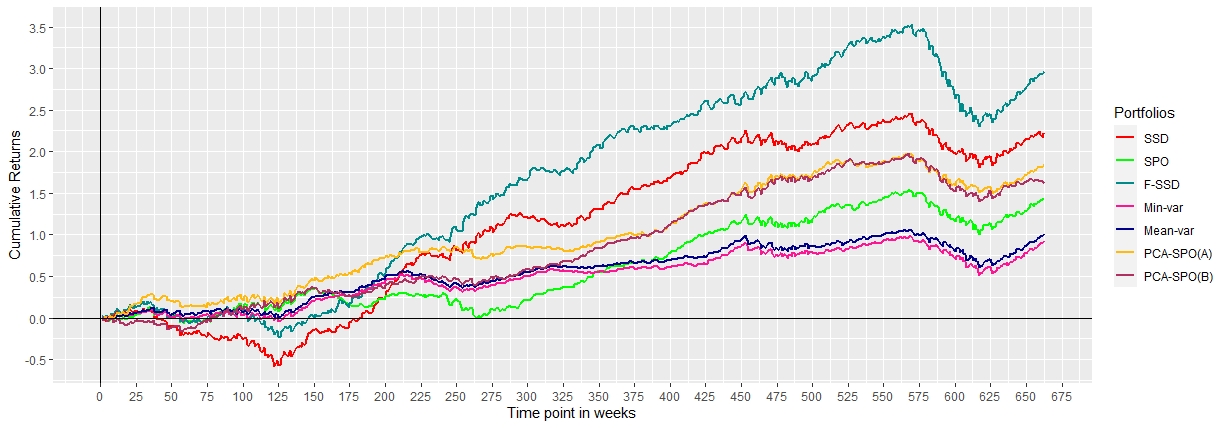}
  \caption{Cumulative returns of the portfolios from PCA-SPO(A), PCA-SPO(B), SPO, SSD, F-SSD, Min-var and Mean-var for S\&P 500 Market (U.S.) for 3 month out-of-sample period (rolling window scheme)}
  \label{fig:CR_US}
\end{figure}

\begin{figure}[!ht]
  \includegraphics[height=6cm, width=\textwidth]{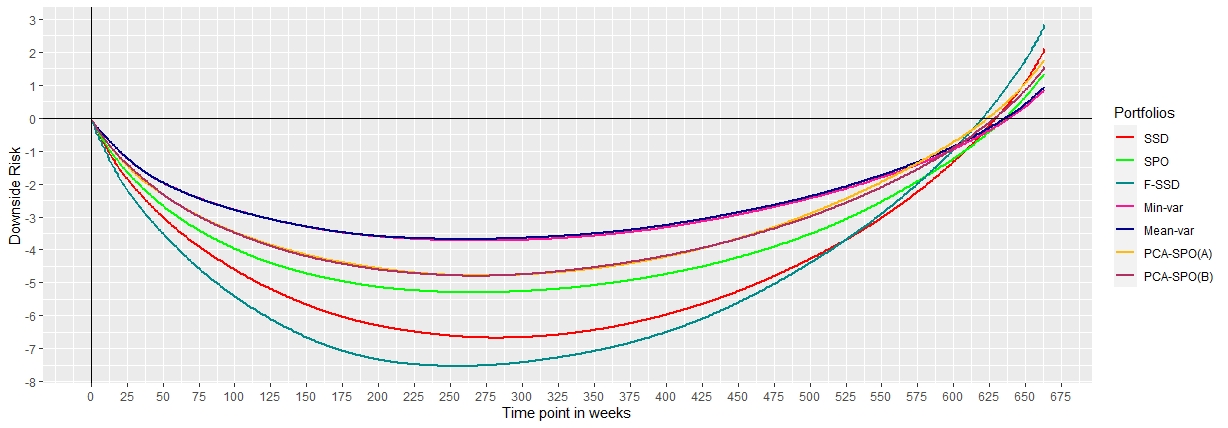}
  \caption{Downside Risk of the portfolios from PCA-SPO(A), PCA-SPO(B), SPO, SSD, F-SSD, Min-var and Mean-var for S\&P 500 Market (U.S.) for 3 month out-of-sample period (rolling window scheme)}
  \label{fig:DR_US}
\end{figure}

Figures \ref{fig:CR_IND} and \ref{fig:CR_US} depict the out-of-sample cumulative returns of all the portfolios over the data sets, S\&P BSE 500 and S\&P 500, respectively. The main points of observations recorded from these figures are as follows:

\begin{enumerate}
    \item From the Figure \ref{fig:CR_IND}, portfolios from SSD dominates all till 130th week, followed by PCA-SPO(B). While portfolios from PCA-SPO(A) clocks the least return for first 130 weeks, it however increases thereafter and dominates all other portfolios from the 225th week. Portfolios from F-SSD suffers with the most negative returns and spots the least returns from 150th week to the 415th week.  
    
    \item From the Figure \ref{fig:CR_US}, the portfolios from PCA-SPO(A) dominates all portfolios till 210th week, while SSD suffers with the least returns (most negative returns) during this phase. SPO, on the other hand suffers with the least return from 200th week to 340th week. Portfolios from SSD, F-SSD, PCA-SPO(A), PCA-SPO(B) and SPO observe a rise (with F-SSD dominating) from 325th week till 575th week during the bullish phase of the market, indicating the dominance of expected returns over FRs. However, both SSD and F-SSD also observe a sharp fall during the bearish phase from 575th week to 615th week, with a drop of 1.25 cumulative returns, while the portfolios from remaining models drop only by 0.6.

    \item We further note that the portfolios from PCA-SPO(B) never achieve least return over the considered time horizon for both the markets indicating the safer portfolio for the loss averse investors. In comparison to the SPO, PCA-SPO(A) always outperforms it, while PCA-SPO(B) outperforms SPO for S\&P 500 and depicts similar return movement as the SPO over S\&P BSE 500, highlighting the importance of selection of dominant ratios using PCA. 
    
\end{enumerate}

Figures \ref{fig:DR_IND} and \ref{fig:DR_US} depicts the 
downside risk pattern\footnote{Let $r_{i}^{k}$ denote the portfolio return in the $i$th window for model/strategy $k$. To trace the downside risk, we first sort the return series of portfolio $z$ in ascending order as $r_{z1}^{k}<r_{z2}^{k}< \cdots <r_{zM}^{k}$, where $M$ denotes the number of rolling windows generated and then calculate the cumulative returns of the sorted return series. The sorted cumulative return series for each portfolio is then plotted in R software. Note that lower the graph, higher is the downside risk from the respective portfolio.} from all the models over the S\&P BSE 500 (India) and S\&P 500 (U.S.) markets, respectively. We record the following main points from these figures:

\begin{enumerate}
    \item We observe that the downside risk graphs corresponding to the models F-SSD and SSD are falling below in comparison to all other models for both the markets, indicating higher downside risk from these models, while those from the models Min-var and Mean-var stand above all other models (for both the markets) suggesting lowest downside risk from these models. 
    
    \item The downside risk graphs from the models SPO, PCA-SPO(A), and PCA-SPO(B) are found hanging above than the models SSD and F-SSD, highlighting the relevance of FRs and the two-step sector-wise implementation over the one-step F-SSD model. 
    
    \item Further, the better downside risk graphs from PCA-SPO(A) and PCA-SPO(B) in comparison to SPO highlights the relevance of selecting critical ratios for the sectors when FRs are considered. 
\end{enumerate}

To summarize, the proposed strategies PCA-SPO(A) and PCA-SPO(B) demonstrates overall best performance in terms of both risk as well return-risk measures. While SSD and F-SSD outperforms in terms of return, they however prove to be the most risky. On the other hand, Mean-var and Min-var depict better performance in terms of risk, however fail to impress in terms of return and performance ratios. Thus, the results from the rolling window scheme validates the proposed strategies PCA-SPO(A) and PCA-SPO(B), highlighting the importance of FRs, their sector-wise selection using PCA and the two-step implementation in the portfolio selection process.

\subsubsection{Market phase wise out-of-sample analysis}

\begin{figure}[!ht]
  \includegraphics[height=5cm, width=\textwidth]{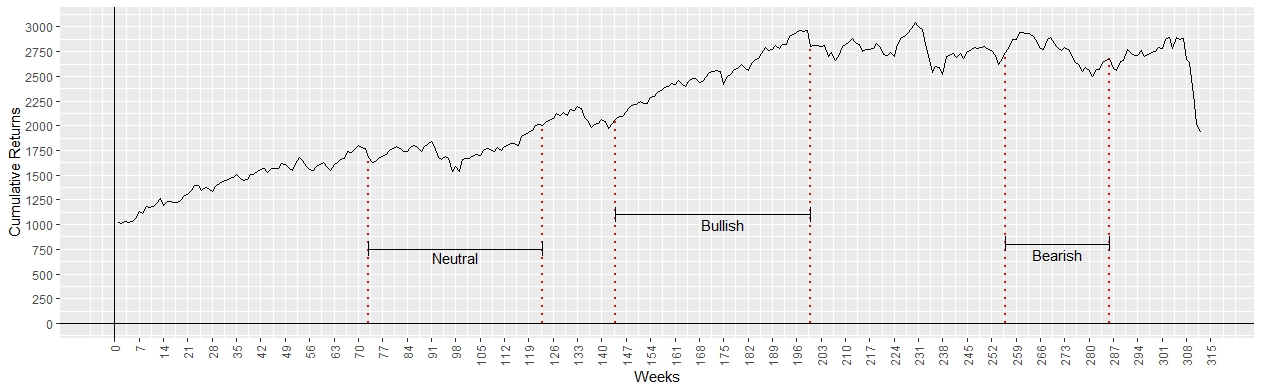}
  \caption{Cumulative Returns for S\&P BSE 500 Index (India)}
  \label{fig:SC_IND}
\end{figure}

\begin{figure}[!ht]
  \includegraphics[height=5cm, width=\textwidth]{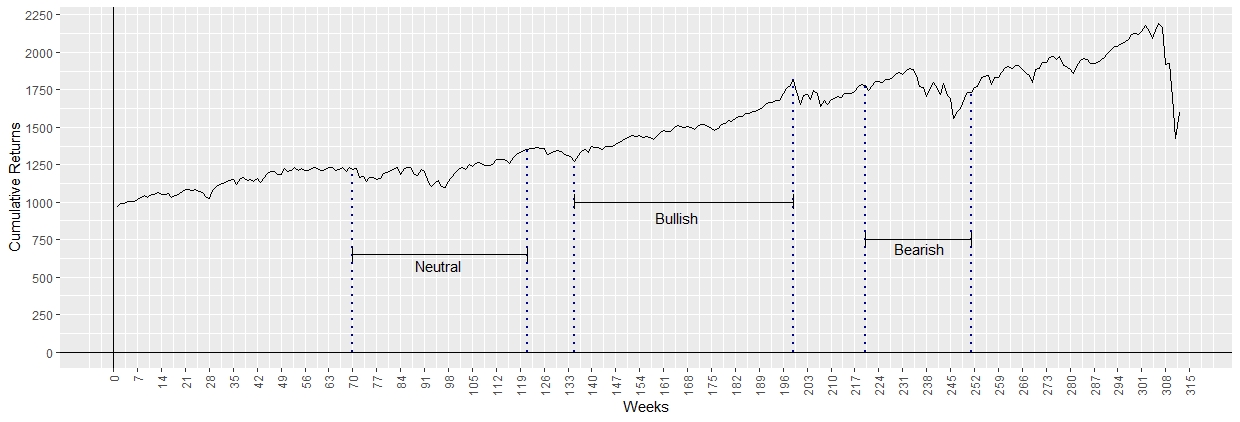}
  \caption{Cumulative Returns for S\&P 500 Index (U.S.)}
  \label{fig:SC_US}
\end{figure}

For a comprehensive analysis of the proposed strategies with the existing models, we conduct the out-of-sample analysis on three different phases (price movements) of the market: neutral, bullish and bearish. For this, we first plot the cumulative returns graph for each of the index and identify cuts corresponding to the three phases graphically. Figures \ref{fig:SC_IND} and \ref{fig:SC_US} show the labels for different market phases for the indices S\&P BSE 500 (India) and S\&P 500 (U.S.) respectively over the time horizon of 6 years from April 2014 to March 2020.

Each of the labelled scenario (neutral, bullish and bearish) is observed as out-of-sample period (testing period) and the corresponding previous year (52 weeks) is considered as the in-sample period (training period) for both the indices. All models are solved over the in-sample period and the portfolios thus obtained are tested over the identified phases on the basis of the performance measures mentioned in Section \ref{sec:5.1}.

\begin{table}[!ht]
    \centering
    \caption{Out-of-sample performance analysis of optimal portfolios generated from the model for neutral market scenario}
    \footnotesize
    \label{tab:neutral}
     \begin{tabular*}{\textwidth}{@{\extracolsep{\fill}}lccccccc@{\extracolsep{\fill}}}
\noalign{\smallskip}\hline\noalign{\smallskip}
Portfolios & SSD & SPO & Min-Var & Mean-Var &  PCA-SPO(A) & PCA-SPO(B) & F-SSD  \\
\noalign{\smallskip}\hline\noalign{\smallskip}
\multicolumn{8}{c}{S\&P BSE 500}\\
\noalign{\smallskip}\hline
Mean Return & 0.00033 & 0.00182 & 0.00153 & $-0.00240$ & 0.00200 & 0.00248 & $-0.00029$ \\
Sharpe Ratio & 0.00941 & 0.06793 & 0.07917 & \textbf{-----} & 0.07225 & 0.09047 & \textbf{-----}\\
Sortino Ratio & 0.01346 & 0.10104 & 0.10954 & \textbf{-----} & 0.09972 & 0.12687 & 0.00000\\
DD & 0.02457 & 0.01802 & 0.01399 & 0.02985 & 0.02003 & 0.01954  & 0.02156 \\
VaR$_{0.95}$ & 0.06096  & 0.04029 & 0.03067  & 0.06121 & 0.03685 &  0.03477 & 0.04896 \\
VaR$_{0.97}$ & 0.06407 & 0.04574 & 0.04188 & 0.08558 & 0.07083 & 0.06414 & 0.06068 \\
CVaR$_{0.95}$ & 0.07696 & 0.05001 & 0.03858 & 0.08876 & 0.06221 & 0.05780  & 0.06807\\
CVaR$_{0.97}$ & 0.08496 & 0.05487 & 0.04254 & 0.010253 & 0.07489 & 0.06931 & 0.07763\\
STARR$_{0.95}$ & 0.00430 & 0.03640 & 0.03972 & \textbf{-----} & 0.03211 & 0.04290 & \textbf{-----} \\
STARR$_{0.97}$ & 0.00389 & 0.03318 & 0.03602 & \textbf{-----} & 0.02667 & 0.03577 & \textbf{-----} \\
Rachev$_{0.95}$ & 0.87265 & 1.20308 & 0.99457 & 0.97704 & 0.93078 & 0.97278 & 0.70815\\
Rachev$_{0.97}$ & 0.80938 & 1.24301 & 1.01783 & 0.91585 & 0.85679 & 0.91257 & 0.63054\\
\noalign{\smallskip}\hline
\multicolumn{8}{c}{S\&P 500}\\
\noalign{\smallskip}\hline
Mean Return & 0.00001 & 0.00195 & 0.00191 & 0.00142 & 0.00356 & 0.00309 & -0.00078 \\
Sharpe Ratio & 0.00046 & 0.06844 & 0.11154 & 0.04126 & 0.13519 & 0.11029 & 0.00000\\
Sortino Ratio & 0.00063 & 0.09048 & 0.14412 & 0.06224 & 0.17136 & 0.14108 & 0.00000\\
DD & 0.02108 & 0.02157 & 0.01327 & 0.02286 & 0.02079 & 0.02188  & 0.01851 \\
VaR$_{0.95}$ & 0.05758  & 0.06328 & 0.03606  & 0.05066 & 0.04922 &  0.04447 & 0.04524 \\
VaR$_{0.97}$ & 0.06245 & 0.06863 & 0.04104 & 0.07241 & 0.06057 & 0.04689 & 0.04932 \\
CVaR$_{0.95}$ & 0.06355 & 0.06773 & 0.04082 & 0.06843 & 0.06715 & 0.06410  & 0.06428\\
CVaR$_{0.97}$ & 0.06654 & 0.06996 & 0.04319 & 0.07731 & 0.07611 & 0.07392 & 0.07814\\
STARR$_{0.95}$ & 0.00021 & 0.02882 & 0.04685 & 0.02079 & 0.05304 & 0.04816 & 0.00000\\
STARR$_{0.97}$ & 0.00020 & 0.02790 & 0.04427 & 0.01840 & 0.04680 & 0.04176 & 0.00000\\
Rachev$_{0.95}$ & 0.79001 & 0.74628 & 0.69671 & 1.22702 & 0.66233 & 0.63831 & 0.82042\\
Rachev$_{0.97}$ & 0.78460 & 0.80210 & 0.69115 & 1.31646 & 0.61109 & 0.56506 & 0.80262\\
\noalign{\smallskip}\hline
\end{tabular*}
\end{table}

Table \ref{tab:neutral} records the performance of the portfolios for \textit{neutral} out-of-sample phase for the indices S\&P BSE-500 (India) and S\&P 500 (U.S.). Following observations are drawn from Table \ref{tab:neutral}:

\begin{enumerate}
    \item For the Indian S\&P BSE 500 index, PCA-SPO(B) performs best in terms of mean return, Sharpe, Sortino and STARR$_{0.95}$ values, while for the S\&P 500 index, PCA-SPO(A) outperforms all other models in terms of mean return and the three ratios. 
    
    \item Both the PCA-SPO(A) and PCA-SPO(B) outperform SSD and SPO in terms of mean return, Sharpe, Sortino, VaR$_{0.95}$, VaR$_{0.97}$, CVaR$_{0.95}$ and STARR values for both the indices, highlighting the importance of FRs and the relevance of selecting critical ratios for each sector using PCA. 
    
    \item Portfolios from the Min-var model performs best in terms of risk measures by recording the least values of DD, VaR and CVaR values for both data sets, but does not impress much in terms of return-risk ratios. 
    
    \item Portfolios from the F-SSD model generates negative mean returns for both the data sets and thus suffers with the lowest values of Sharpe, Sortino and STARR ratios. In terms of risk measures, the portfolios from Mean-var models record the largest values of DD, VaR$_{0.95}$, VaR$_{0.97}$, CVaR$_{0.95}$ and CVaR$_{0.97}$ for S\&P BSE 500 and of DD, VaR$_{0.97}$ and CVaR$_{0.95}$ for S\&P 500 data set.  
    
    \item Portfolios from SSD and F-SSD also exhibit poor performance in terms of risk measures as well as the return-risk ratios, viz., Sharpe, Sortino, STARR and Rachev ratios. 
    
    \item None of the proposed strategies PCA-SPO(A) or PCA-SPO(B) perform worst in terms of any of the risk measure, making them safe portfolios over the neutral phase of the market. 
\end{enumerate}

In a nutshell, both proposed strategies PCA-SPO(A) and PCA-SPO(B) prove favourable during the neutral phase of the market.

\begin{table}[!ht]
    \centering
    \caption{Out-of-sample performance analysis of optimal portfolios generated from the model for bearish market scenario}
    \footnotesize
    \label{tab:bear}
    \begin{tabular*}{\textwidth}{@{\extracolsep{\fill}}lccccccc@{\extracolsep{\fill}}}
\noalign{\smallskip}\hline\noalign{\smallskip}
Portfolios & SSD & SPO & Min-Var & Mean-Var &  PCA-SPO(A) & PCA-SPO(B) & F-SSD  \\
\noalign{\smallskip}\hline\noalign{\smallskip}
\multicolumn{8}{c}{S\&P BSE 500}\\
\noalign{\smallskip}\hline
Mean Return & 0.00471 & 0.00503 & 0.00222 & 0.00878 & 0.00159 & 0.00426 & 0.00607 \\
Sharpe Ratio & 0.20848 & 0.27304 & 0.11545 & 0.24921 & 0.08654 & 0.246720 & 0.28483\\
Sortino Ratio & 0.30656 & 0.38065 & 0.16069 & 0.41441 & 0.12028 & 0.36041 & 0.38619\\
DD & 0.01535 & 0.01322 & 0.01381 & 0.02119 & 0.01321 & 0.01182 & 0.01572\\
VaR$_{0.95}$ & 0.02299  & 0.02188 & 0.02014  & 0.03748 & 0.02719 &  0.02051 & 0.03591 \\
VaR$_{0.97}$ & 0.03083 & 0.02384 & 0.02440 & 0.04053 & 0.03082 & 0.02545 & 0.04745 \\
CVaR$_{0.95}$ & 0.03252 & 0.02683 & 0.03232 & 0.04294 & 0.03057 & 0.02390  & 0.05378\\
CVaR$_{0.97}$ & 0.03728 & 0.02931 & 0.03841 & 0.04567 & 0.03225 & 0.02559 & 0.05103\\
STARR$_{0.95}$ & 0.14475 & 0.18758 & 0.06866 & 0.20451 & 0.05197 & 0.17827 & 0.11291\\
STARR$_{0.97}$ & 0.12625 & 0.17171 & 0.05777 & 0.19229 & 0.04926 & 0.16646 & 0.11900\\
Rachev$_{0.95}$ & 1.42063 & 1.31205 & 1.07080 & 1.90883 & 0.98161 & 1.45688 & 1.61128\\
Rachev$_{0.97}$ & 1.42651 & 1.39050 & 0.95888 & 2.03490 & 0.98576 & 1.45545 & 0.95994\\
\noalign{\smallskip}\hline\noalign{\smallskip}
\multicolumn{8}{c}{S\&P 500}\\
\noalign{\smallskip}\hline\noalign{\smallskip}
Mean Return & $-0.00590$ & $-0.00187$ & $-0.00285$ & $-0.01191$ & $-0.00420$ & $-0.00418$ & $-0.00658$ \\
Sharpe Ratio & \textbf{-----} & \textbf{-----} & \textbf{-----} & \textbf{-----} & \textbf{-----} & \textbf{-----} & \textbf{-----}\\
Sortino Ratio & \textbf{-----} & \textbf{-----} & \textbf{-----} & \textbf{-----} & \textbf{-----} & \textbf{-----} & \textbf{-----}\\
DD & 0.02800 & 0.02562 & 0.01823 & 0.05565 & 0.02273 & 0.02262  & 0.03891 \\
VaR$_{0.95}$ & 0.07243  & 0.06543 & 0.05112  & 0.10361 & 0.06344 &  0.05930 & 0.09534 \\
VaR$_{0.97}$ & 0.07808 & 0.06766 & 0.06882 & 0.26427 & 0.06944 & 0.06402 & 0.13045 \\
CVaR$_{0.95}$ & 0.09406 & 0.08318 & 0.07496 & 0.22992 & 0.08305 & 0.07708  & 0.14112\\
CVaR$_{0.97}$ & 0.08133 & 0.07048 & 0.07168 & 0.27528 & 0.07233 & 0.06669 & 0.13589\\
STARR$_{0.95}$ & \textbf{-----} & \textbf{-----} & \textbf{-----} & \textbf{-----} & \textbf{-----} & \textbf{-----} & \textbf{-----}\\
STARR$_{0.97}$ & \textbf{-----} & \textbf{-----} & \textbf{-----} & \textbf{-----} & \textbf{-----} & \textbf{-----} & \textbf{-----}\\
Rachev$_{0.95}$ & 0.71782 & 0.79315 & 0.56139 & 0.46747 & 0.80237 & 0.82596 & 0.69825\\
Rachev$_{0.97}$ & 0.78877 & 0.79163 & 0.52777 & 0.41687 & 1.05939 & 0.88585 & 0.68115\\
\noalign{\smallskip}\hline
\end{tabular*}
\end{table}

Table \ref{tab:bear} tracks down the performance of the portfolios for \textit{bearish} out-of-sample phase for the indices S\&P BSE-500 (India) and S\&P 500 (U.S.). We draw the following observations from Table \ref{tab:bear}:

\begin{enumerate}
    \item For the S\&P BSE 500 dataset, though the Mean-var model exhibits better performance in terms of mean return, Sortino, STARR and Rachev ratios but  records the highest values of DD and VaR$_{0.95}$ and second highest values of VaR$_{0.97}$, CVaR$_{0.95}$ and CVaR$_{0.97}$ after the F-SSD model. 
    For the S\&P 500 data set, it records the most negative mean return and highest values of DD, VaR$_{0.95}$, VaR$_{0.97}$, CVaR$_{0.95}$ and CVaR$_{0.97}$, proving to be the riskiest of all models considered along with the F-SSD and SSD model respectively, stand as the second and third most riskiest models. This hints that not using FRs at all or analyzing all stocks together on the basis of FRs may mislead the investor and result in loss during the bearish scenario.  
    
    \item PCA-SPO(B) records the lowest values of DD, CVaR$_{0.95}$, CVaR$_{0.97}$ and second lowest value of VaR$_{0.95}$ for S\&P BSE 500, while for S\&P 500, it records the lowest values of VaR$_{0.97}$ and CVaR$_{0.97}$ and second lowest values of DD and VaR$_{0.95}$, proving advantageous in terms of risk in a bearish scenario. Whereas, PCA-SPO(A) performs decently in terms of risk, it records the worst values of mean return, Sharpe, Sortino, STARR and Rachev ratios for S\&P BSE 500, under performing in comparison to SPO and PCA-SPO(B), which hints the importance of each category of ratio when FRs are considered in portfolio selection. 
    
    \item Though the model SPO achieves higher mean return, Sharpe and Sortino values than PCA-SPO(B) but the later performs consistently well in terms of risk and Rachev ratios for both data sets. This hints the importance of each category of ratio and how choosing critical ratio using PCA can prove advantageous in terms of risk during a bearish scenario, where one aims to be risk-averse. 
    
    \item While the proposed strategy PCA-SPO(B) never generates worst values for any performance measures for both the data sets, PCA-SPO(A) fails to impress in bearish scenario, particularly for the Indian data set by recording worst values of mean return, Sharpe, Sortino, STARR and Rachev ratios.
    
    \end{enumerate}

To sum up, for a bearish scenario, where the primary aim of any investor is to protect the portfolio from extreme losses, the proposed strategy PCA-SPO(B) proves to be the most advantageous by winning over risk criteria and at the same time not generating poor returns. 

\begin{table}[htp!]
    \centering
    \caption{Out-of-sample performance analysis of optimal portfolios generated from the model for bullish market scenario}
    \footnotesize
    \label{tab:bull}
     \begin{tabular*}{\textwidth}{@{\extracolsep{\fill}}lccccccc@{\extracolsep{\fill}}}
\noalign{\smallskip}\hline\noalign{\smallskip}
Portfolios & SSD & SPO & Min-Var & Mean-Var &  PCA-SPO(A) & PCA-SPO(B) & F-SSD  \\
\noalign{\smallskip}\hline\noalign{\smallskip}
\multicolumn{8}{c}{S\&P BSE 500}\\
\noalign{\smallskip}\hline
Mean Return & 0.00831 & 0.00434 & 0.00298 & 0.00941 & 0.00965 & 0.00481 & 0.00807 \\
Sharpe Ratio & 0.28451 & 0.22015 & 0.27989 & 0.21523 & 0.37824 & 0.24348 & 0.32661\\
Sortino Ratio & 0.40129 & 0.31834 & 0.46411 & 0.33365 & 0.56663 & 0.35434 & 0.49740\\
DD & 0.02070 & 0.01362 & 0.00642 & 0.02821 & 0.01703 & 0.01357  & 0.01623 \\
VaR$_{0.95}$ & 0.03601  & 0.02962 & 0.01613  & 0.07264 & 0.02964 &  0.02155 & 0.03235 \\
VaR$_{0.97}$ & 0.06381 & 0.04153 & 0.02253 & 0.09424 & 0.03719 & 0.03715 & 0.03460 \\
CVaR$_{0.95}$ & 0.05544 & 0.04102 & 0.02188 & 0.09478 & 0.03593 & 0.03639  & 0.03977\\
CVaR$_{0.97}$ & 0.06516 & 0.04673 & 0.02475 & 0.10584 & 0.03907 & 0.04381 & 0.04736\\
STARR$_{0.95}$ & 0.14983 & 0.10571 & 0.13623 & 0.09930 & 0.26858 & 0.13215 & 0.20299\\
STARR$_{0.97}$ & 0.12749 & 0.09280 & 0.12042 & 0.08892 & 0.26858 & 0.13215 & 0.20299\\
Rachev$_{0.95}$ & 1.12171 & 1.04961 & 1.04004 & 1.13929 & 1.89863 & 1.21669 & 1.67022\\
Rachev$_{0.97}$ & 0.98077 & 0.96266 & 0.93124 & 1.09607 & 1.88073 & 1.03221 & 1.60978\\
\noalign{\smallskip}\hline\noalign{\smallskip}
\multicolumn{8}{c}{S\&P 500}\\
\noalign{\smallskip}\hline
Mean Return & 0.01156 & 0.00607 & 0.00475 & 0.02110 & 0.00586 & 0.00840 & 0.01227 \\
Sharpe Ratio & 0.47199 & 0.37377 & 0.49220 & 0.32891 & 0.35436 & 0.45482 & 0.44257\\
Sortino Ratio & 0.69521 & 0.57807 & 0.68732 & 0.57574 & 0.53203 & 0.66145 & 0.06743\\
DD & 0.01663 & 0.01050 & 0.00692 & 0.03664 & 0.01102 & 0.01269  & 0.01819 \\
VaR$_{0.95}$ & 0.03179  & 0.02591 & 0.00986  & 0.06728 & 0.02257 &  0.02158 & 0.03854 \\
VaR$_{0.97}$ & 0.03819 & 0.02679 & 0.01926 & 0.07895 & 0.02998 & 0.02928 & 0.04217 \\
CVaR$_{0.95}$ & 0.03391 & 0.02490 & 0.01500 & 0.07234 & 0.02591 & 0.02786  & 0.03873\\
CVaR$_{0.97}$ & 0.04021 & 0.02820 & 0.01994 & 0.08606 & 0.03160 & 0.03537 & 0.04478\\
STARR$_{0.95}$ & 0.34108 & 0.24388 & 0.31689 & 0.29165 & 0.22631 & 0.30132 & 0.31676\\
STARR$_{0.97}$ & 0.28761 & 0.21528 & 0.23835 & 0.24514 & 0.18552 & 0.23736 & 0.27395\\
Rachev$_{0.95}$ & 1.72434 & 1.75461 & 1.39755 & 2.79556 & 1.54505 & 1.70714 & 2.04578\\
Rachev$_{0.97}$ & 1.71179 & 1.84416 & 1.19978 & 3.17779 & 1.52706 & 1.63193 & 2.06018\\
\noalign{\smallskip}\hline
\end{tabular*}
\end{table}

Table \ref{tab:bull} documents the performance of the portfolios for \textit{bullish} out-of-sample market scenario and the observations can be summarized as follows:

\begin{enumerate}
    \item For the Indian S\&P BSE 500 market, PCA-SPO(A) depicts superior performance in terms of mean return, Sharpe, Sortino, STARR and Rachev ratios but fails to perform the same over S\&P 500 market. 
    
    \item The portfolios from Mean-var model outperforms in terms of mean return for both the data sets but proves to be the most risky by recording highest values of DD, VaR$_{0.95}$, VaR$_{0.97}$, CVaR$_{0.95}$ and CVaR$_{0.97}$ for both the data sets. In terms of risk, the Min-var model proves best; however achieves the worst values of mean return and Rachev ratios for both the data sets.
    
    \item The portfolios from the F-SSD and SSD model achieves higher values of mean return, Sharpe and Sortino ratios in comparison to the portfolios from PCA-SPO(B); however also prove riskier than PCA-SPO(B) by recording higher values of DD, VaR$_{0.95}$, VaR$_{0.97}$, CVaR$_{0.95}$ and CVaR$_{0.97}$ with SSD recording the highest values over Indian data set and F-SSD over the U.S. data set.

    \item  The proposed strategy PCA-SPO(B) outperforms SPO for almost all performance measures and never achieves worst performance for any performance measure for both the data sets. 
    
\end{enumerate}

To summarize, for a bullish scenario, the aggressive (risk-taking) investors should opt for mean-variance or SSD or the F-SSD (when FRs are to be considered) whereas conservative (risk-averse) should opt for the minimum variance model. For the moderate investors however, PCA-SPO(B) proves to be the most advantageous that wisely employs financial analysis to avoid extreme risk and at the same time benefits from the rising return trend in a bullish scenario.

\section{Managerial Implications}

Our proposed strategy first filters significant objective functions made of financial ratios for each sector via PCA and then apply an optimization tools to find out the optimal weights for each stock and thus helping portfolio managers and investors with a sophisticated portfolio selection strategy. The applications of this study, however, is not just limited to portfolio management but has important implementations for accountants, business analysts, financial analysts, researchers and academicians. The proposed strategy can be exploited as a whole by a portfolio analyst who is focusing on a diversified investment across all sectors by utilizing the fundamental analysis with the help of financial ratios in the investment process. Nonetheless, each step of the strategy has an individual applicability in various domains. For instance, Step 1 that uses PCA to filter out relevant ratios can be used by practitioners for determining the most significant factors from a large set of variables for various purposes like assessment of risk, forecasting failures, predicting future profits, and many more. Additionally, Step 2 that determines an optimal portfolio for each sector can be utilized by portfolio managers and financial analysts to filter out elite stocks from each sector. Analogous to this, banks can utilize this step for loan assessment. This step can help banking analysts to filter out A-list businesses to grant loan to. Thus, the proposed research has managerial implications for portfolio management, forecasting and predictive analysis, banking and education.

\section{Conclusion}
\label{sec: 6}
The importance of FA in portfolio framework can not be overlooked. However, the right choice of ratios and their correct usage is even more crucial, owing to the fact that companies belonging to different industry sectors have different operational structures and business functionalities. 

The SPO model aptly considers the sector-wise implementation but chooses ratios common for all sectors. This paper is an attempt forward on improvising the SPO model by implementing PCA for the selection of critical ratios for each sector, extracted in two ways, one from the components (strategy named as PCA-SPO(A)) and other from each category of ratios on the basis of communalities (strategy named as PCA-SPO(B)). We conduct the comparative analysis on the proposed strategies, PCA-SPO(A) and PCA-SPO(B), with the models Mean-var, Min-var, F-SSD, SSD, and SPO in two ways, viz., a rolling window scheme and second using scenario based analysis.

The rolling window analysis which is carried over the out-of-sample of 3 months showed that the optimal portfolios obtained from the proposed strategies PCA-SPO(A) and PCA-SPO(B) resulted in lower values of risk measures and generated better risk-reward profile in terms of Sortino, Sharpe, Rachev, and STARR ratios in comparison to other models for both the data sets considered in the study. On the other hand, the scenario based analysis which is carried over three different phases of the market, viz., neutral, bullish and bearish endorses the proposed strategy PCA-SPO(B). While for the bullish scenario, PCA-SPO(B) proves promising for the moderate investors only, it however proves propitious for all types of investors for both neutral and bearish scenarios.

To sum up, the proposed strategy PCA-SPO(B) is an apt decision making process that focuses primarily on the extraction of critical ratios from each category by employing PCA sector wise and then generates final portfolio incorporating the two step SPO model. The financial benefits of PCA-SPO(B) over SPO and other models confirms the practical usage of the proposed strategy. Further extensions of the present work are possible by including other important factors like technical indicators, sustainability, socio-economic variables, and transaction costs in the portfolio framework.


\end{document}